\documentclass[twoside]{article}

\usepackage[accepted]{aistats2026}
\usepackage[round]{natbib}
\newcommand{\indep}{\mathrel{\text{\scalebox{1.07}{$\perp\mkern-10mu\perp$}}}}

\usepackage{xcolor}
\usepackage{amssymb}
\usepackage{multirow}
\usepackage{booktabs}
\usepackage{graphicx}
\usepackage{subcaption}
\usepackage{hyperref}
\usepackage{cleveref}
\usepackage{algorithm}
\usepackage{comment}

\usepackage{amsmath}    
\usepackage{amssymb}    
\usepackage{amsfonts}   
\usepackage{amsthm}     
\usepackage{mathtools}  
\usepackage{bm}         
\usepackage{mathrsfs}   

\usepackage{enumitem}   
\usepackage{hyperref}   

\theoremstyle{plain}
\newtheorem{theorem}{Theorem}

\begin{document}

%

%

\twocolumn[

\aistatstitle{Empirically Calibrated Conditional Independence Tests}

\aistatsauthor{ Milleno Pan \And Antoine de Mathelin \And  Wesley Tansey }

\aistatsaddress{Computational Oncology \\ Memorial Sloan Kettering Cancer Center} ]


\begin{abstract}
Conditional independence tests (CIT) are widely used for causal discovery and feature selection. Even with false discovery rate (FDR) control procedures, they often fail to provide frequentist guarantees in practice. We highlight two common failure modes: (i) in small samples, asymptotic guarantees for many CITs can be inaccurate and even correctly specified models fail to estimate the noise levels and control the error, and (ii) when sample sizes are large but models are misspecified, unaccounted dependencies skew the test’s behavior and fail to return uniform p-values under the null. We propose Empirically Calibrated Conditional Independence Tests (ECCIT), a method that measures and corrects for miscalibration. For a chosen base CIT (e.g., GCM, HRT), ECCIT optimizes an adversary that selects features and response functions to maximize a miscalibration metric. ECCIT then fits a monotone calibration map that adjusts the base-test p-values in proportion to the observed miscalibration. Across empirical benchmarks on synthetic and real data, ECCIT achieves valid FDR with higher power than existing calibration strategies while remaining test agnostic. Code is available at \href{https://github.com/tansey-lab/ECCIT}{https://github.com/tansey-lab/ECCIT}.
\end{abstract}


\section{INTRODUCTION}

The central tool for rigorously detecting causal relationships in the presence of confounders is the conditional independence test.
Mathematically, there exists a causal edge if two variables $X$ and $Y$ are dependent after controlling for all confounders $Z$; thus, the null hypothesis of no causal effect is one of conditional independence,
\begin{equation}
\label{eq:cond_indep}
    \mathcal{H}_0 \colon X \indep Y \mid Z \, .
\end{equation}
One common use case for conditional independence testing is the controlled variable selection problem. Given data $\{ (X_1, X_2, \ldots, X_m, Y)_i \}_{i=1}^{n}$, we wish to find the set of variables $S \subseteq [m]$ such that $X_j \indep Y \mid X_{-j}$ if and only if $j \not \in S$. That is, we want the Markov blanket of $Y$. If there are no latent confounders and all $X_j$ variables are known not to be caused by $Y$ (e.g. if each $X_j$ was observed before $Y$), then $S$ corresponds to the set of causal features of $Y$.

In real world settings with finite data and noisy observations, it is impossible to infer $S$ without some error rate. Here, we work within the frequentist hypothesis testing framework: the user supplies the procedure with an acceptable error rate $\alpha$ and the procedure returns a candidate set $\hat{S}$. Valid testing procedures provide a guarantee that the expected error on $\hat{S}$ will be no larger than $\alpha$, the user-specified threshold. After controlling the error rate, procedures can be compared based on whether one has a higher true positive rate, also known as power.

Statistical methods for testing \cref{eq:cond_indep} face a challenging task. Theoretically, it is impossible to produce a method capable of having non-trivial power against all possible alternative hypotheses~\citep{shah:peters:2020:hardness-conditional-independence}. Empirically, many methods struggle to control the error rate when the number of features $m$ is large relative to the sample size $n$. Nonparametric methods converge too slowly to produce valid frequentist $p$-values. Parametric assumptions on the structure of the possible conditional dependencies improves sample efficiency, but open one up to model misspecification. Addressing the issue of error rate control in the high-dimensional and low-sample regimes remains an open problem, motivating a growing literature on different approaches \citep{tansey:etal:2021:hrt,sudarshan:etal:2021:contra-cit,li:etal:2023:maxway-crt}.

The importance of the problem is highlighted by the growing popularity of conditional independence tests in science. CITs have been applied to a wide array of biology and medical data~\citep{shen:etal:2019:cancer-biomarker-knockoffs,bates:etal:2020:genetic-trio-knockoffs,tansey:etal:2021:dose-response,barry:etal:2021:sceptre, sudarshan:etal:2020:ddlk,niu:etal:2024:spacrt-crispr} and used in development of models that have been integrated into electronic health records \citep{razavian2020validated}. The ground truth error rate of these procedures is unknowable, but it is common for scientific datasets to fall into the high-dimensional or low-sample regimes where current CIT procedures typically fail.

In this paper, we propose Empircally Calibrated CITs (ECCITs) as a wrapper method for a broad class of conditional independence testing procedures. ECCITs take an existing conditional independence testing or controlled variable selection procedure, paired with a dataset on which a user wishes to apply it. The empirical calibration method then optimizes an adversarial model to maximally inflate the error rate of the CIT procedure on the target dataset. The $p$-values from the CIT procedure applied to the dataset are then calibrated such that they would be valid $p$-values even against the adversary. Since ECCITs calibrate against a worst-case function in a class $\mathcal{F}$, the resulting procedure is calibrated or conservative for all other functions in $\mathcal{F}$.

We explore several design choices involved in ECCITs and evaluate their tradeoffs in extensive benchmarks. We evaluate two different types of conditional independence testing procedures, conditional randomization tests and the generalized covariance metric. We consider two different optimization metrics for the adversary and how each choice affects power. We also evaluate several possible function classes for the predictive models used in our uncalibrated CITs, the ground truth function class generating $Y$ from $X$, and the adversary function class against which we are calibrating. In semi-synthetic experiments on a large gene expression dataset, ECCITs outperform a state-of-the-art method for improving robustness of CITs.




\section{BACKGROUND}
\label{sec:background}
Discovering causal relationships equates statistically to discovering conditioning sets that render variables independent. Many methods exist for discovering causal relationships using conditional independence tests~\cite[e.g.][]{spirtes:etal:2000:pc-algorithm,kalisch:buhlmann:2007:high-dimensional-pc,pellet:elisseeff:2008:TC-algorithm,kalisch:buhlmann:2008:robust-pc-algorithm,strobl:etal:2016:pc-p-algorithm}. \citet{kim:etal:2021:local-permutation-tests} divide conditional independence testing methodology into three groups: local permutations, model-X, and asymptotics.
Local permutation methods~\citep{margaritis:2005:distribution-free-bayes-nets,doran:etal:2014:permutation-conditional-independence,sen:etal:2017:model-powered-cond-indep,kim:etal:2021:local-permutation-tests} divide the effect variable into subclasses and test by permuting within each subclass. Subclassing requires either multiple observations of the same confounders or some smoothness assumptions. However, in the case of continuous variables, every observation is almost surely going to be unique. Alternatively, smoothness assumptions (e.g. fixed-width binning as in~\citet{kim:etal:2021:local-permutation-tests}) can be leveraged in combination with kernel-based methods~\citep{fukumizu:etal:2007:kernel-conditional-dependence} and metrics like maximum mean discrepancy~\citep{gretton:etal:2012:kernel-two-sample-test}. Unfortunately, kernel methods are likely to be underpowered in the high-dimensional setting as all samples are going to be far away from each other unless they lie in a low-dimensional subspace. 

Asymptotic methods derive a limiting distribution for a particular test statistic. 
Classical asymptotic methods~\citep{su:white:2008:hellinger-conditional-independence,huang:etal:2010:conditional-independence-correlation} require linear or quadratic parametric forms for the causal relationships. More recent work has focused on flexible models either via kernel-based tests~\citep{zhang:etal:2011:kernel-independence,wang:etal:2015:conditional-distance,strobl:etal:2019:kernel-cond-indep} or through using black box machine learning methods~\citep{shah:peters:2020:hardness-conditional-independence,zhou:etal:2020:conditional-screening,sudarshan:etal:2023:diet-cit}. Black box methods typically work by estimating $\mathbb{E}[X \mid Z]$ and $\mathbb{E}[Y \mid Z]$, then testing the marginal correlation of the residuals of $X$ and $Y$ after subtracting their predicted conditional means. These methods have the benefit of being rate doubly robust: if either the model for $\mathbb{E}[X \mid Z]$ or the model for $\mathbb{E}[Y \mid Z]$ is correctly specified, and the fitted regressions converge sufficiently fast, then the method will asymptotically control the type I error rate. However, in finite samples or with misspecified models, these methods provide no theoretical guarantees and often fail to control the type I error rate, and in practice, it is unknowable whether we are ever in that regime using black box methods.

Model-X methods~\citep{candes:etal:2018:panning} make no assumptions about the relationship between X and Y. Rather, approaches like knockoffs and conditional randomization tests (CRTs) assume access to a large unlabeled dataset on which to build an accurate model of confounders, in particular modeling $P(X \mid Z)$. Such datasets are often available in scientific domains. For example, large genomic~\citep{cheng:etal:2015:msk-impact}, transcriptomic~\citep{garnett:etal:2012:gdsc,weinstein:etal:2013:tcga}, and epigenomic~\citep{drost:clevers:2018:organoids} tumor, cell line, and organoid databases are available for analysis in biology. Some model-X methods also enjoy the doubly robust property asymptotically~\citep{niu:etal:2024:reconciling-model-x-doubly-robust}. In finite samples and high dimensions, accurate estimation of the conditional distributions is challenging. As with doubly robust asymptotic methods, model-X methods often fail to control type I error in practice.

A number of methods have been proposed to increase the practical robustness of both doubly robust asymptotic and model-X methods. The Corrected Pearson Chi-squared CRT~\citep{xu:etal:2024:covariate-shift-crt} adapts CRTs to be robust to covariate shift in the data. The Maxway CRT~\citep{li:etal:2023:maxway-crt} models both $P(X \mid Z)$ and $P(Y \mid Z)$, gaining theoretical and practical advantages over a basic CRT that only models the $X$ conditional distribution. \citet{zhang:etal:2025:doubly-robust-cit-neural} follow a similar strategy by using generative neural network models to estimate both conditional distributions. Perhaps most relevant to this paper, CONTRA~\citep{sudarshan:etal:2021:contra-cit} uses a mixture of real $X$ data and samples from the estimated $P(X \mid Z)$ to train a predictive model of $Y$ used to perform a CRT. By mixing real and synthetic data, the predictive model is trained to predict using a more realistic version of the covariates that will be sampled in practice, thereby reducing the overall type I error rate. While the above methods all help reduce the overall type I error or false discovery rate (FDR) inflation in CITs with finite samples, none of them aim to produce properly calibrated CITs. The gap to-date is therefore that conducting CITs is somewhat robust using these methods, but it is unclear in practice how to calibrate a CIT to control the target error rate in finite samples.

\section{METHOD}
Consider a dataset $\mathcal{D}$ with $m$ features $(X_1, \ldots, X_m)$ and a response variable $Y$. The goal is to conduct controlled variable selection via conditional independence tests of the form $X_j \indep Y \mid X_{-j}$ for $j=1,\ldots,m$.

Assume for now that a conditional independence testing procedure \(T\) has been given; we will consider two concrete examples later. The requirements on \(T\) are only that it returns \(p\)-values for each feature. The issue we wish to resolve is that in practice it may be impossible to know if \(T\) will truly return valid \(p\)-values, i.e., \(\mathrm{Uniform}(0,1)\) under the null hypothesis. This may be due to small sample sizes, large feature counts, or model misspecification within \(T\). Any, none, or all of these issues may be present, but we will be agnostic to the validity of the test and, if it is invalid, the underlying cause. Fix such a test \(T\) that, given \((X,Y)\), returns \(p\)-values we wish to calibrate. For an adversary class \(\mathcal{F}\) of response generators \(Y=f(X,\epsilon)\) and a calibration metric \(M(\cdot,\alpha)\) with target level \(\alpha\), we define the adversary by Eq.~(2), choosing the generator within \(\mathcal{F}\) that makes the chosen metric as large as possible, yielding the worst-case miscalibration for \(T\) relative to the population distribution of \(X\). The expectation is over the distribution of \(X\); since this distribution is not necessarily known in practice, we approximate it by bootstrap resampling of the observed \(X\). Optimizing the bootstrap estimate produces an empirical optimizer \(f^\ast\) and an associated worst-case metric \(M\!\big(T(X,f^\ast(X,\epsilon)),\alpha\big)\).

At a high-level, our proposed empirical calibration method performs the following steps.

1. An adversarial model (chosen from some class \(\mathcal{F}\)) is optimized to maximize the miscalibration of the test statistics under a given calibration metric $M$. This model is then used to generate adversarial outcomes $\tilde{Y}$.

2. The test $T$ is applied to $\tilde{Y}$ to produce adversarial $p$-values alongside $\gamma$, a vector indicating which of the hypothesis tests should be rejected.

3. A monotonic map, which we call the calibrator, is fit to the adversarial $p$-values so that the miscalibration under $M$ is properly controlled. This map is then applied to the p-values generated by the test $T$ on the real data $Y$.

If \(\mathcal{F}\) and $M$ together capture the desired form of error rate control, then by calibrating against the worst-case adversary, the above will yield $p$-values that are either properly calibrated or conservative.

\Cref{alg:eccit} details the full algorithm in generality. We next detail specific design choices and discuss their motivations and impacts.

\begin{algorithm}[t]
\caption{Empirical calibration (adversarial approach).}
\label{alg:eccit}
\small
\begin{tabular}{@{}p{0.97\linewidth}@{}}
\textbf{Input:} Dataset $(X,Y)$; test $T:(X,Y)\mapsto(p_1,\ldots,p_m)$; adversary class $\mathcal{F}$ with $f\in\mathcal{F}$ and $\tilde Y=f(X,\varepsilon)$; calibration metric $M(\cdot,\alpha)$.\\[3pt]

Fit the worst-case adversary $f^\star$ as in \Cref{eq:adversary}.\\
Let $\mathcal{H}_0=\{j:\gamma_j=0\}$ denote the null features under $f^\star$.\\
\textbf{for} $b=1,\ldots,B$ \textbf{do}\\
\quad Bootstrap $X^{(b)}$ by resampling rows of $X$.\\
\quad Sample $\tilde Y^{(b)}=f^\star(X^{(b)},\varepsilon)$.\\
\quad Compute $p^{(b)}=T(X^{(b)},\tilde Y^{(b)})$.\\
\textbf{end for}\\
Construct $\mathrm{Cal}_\alpha$ from $\{(p^{(b)},\mathcal{H}_0^{(b)})\}_{b=1}^B$ for $M(\cdot,\alpha)$.\\[3pt]
\textbf{Return} the calibrated p-values: \\ $p_{\mathrm{cal}}=\mathrm{Cal}_\alpha(T(X,Y))$.\\
\end{tabular}
\end{algorithm}

\subsection{Adversary optimization}

Given $X$ and a testing procedure $T(X, Y)$ that returns $p$-values
$p_1,\dots,p_m$. Let $\mathcal{F}$ be a class of data-generating mechanisms for the
response $Y$, written $Y=f(X,\varepsilon)$ with $f\in\mathcal{F}$ and noise. Given a calibration metric $\mathcal{M}$ and target error rate level $\alpha$, the ECCIT adversary chooses
\begin{equation}
\label{eq:adversary}
f^\star \;\in\; \arg\max_{f\in\mathcal{F}} \; \mathbb{E}_X\big[\mathcal{M}\big(T(X,f(X,\varepsilon)),\,\alpha\big)\big],
\end{equation}
inflating the calibration metric as much as possible within the bounds of the adversarial class and the data distribution $X$. Intuitively, a well calibrated test should be robust against any potential response, so we want to look for the worst case scenario to calibrate against.

\Cref{eq:adversary} requires access to the population distribution over $X$. This is to prevent flexible adversaries from finding edge cases in a single fixed dataset. In practice, we typically do not have access to this sampling distribution. Instead, we generate an approximate expectation using the bootstrap.

The adversary specifies two sets of parameters. The first set is a binary vector $\gamma$ corresponding to which features $X_j$ will be used to generate the synthetic $Y$ ($\gamma_j = 1$) and which will be null variables ($\gamma_j=0$). The second set is the parameters $\theta$ to the function from the non-null features to the synthetic $Y$. For simplicity, we model the response as an additive errors model in practice, though the choice is flexible. Specifically, let $\tilde{Y}$ be the adversarial response such that,
\[
\tilde{Y} \;=\; \mu_\theta\!\big(X\cdot \gamma\big)\;+\;\varepsilon,\quad
\varepsilon\sim P(\varepsilon),\quad \mathbb{E}[\varepsilon] = 0 \, ,
\]
where $\mu_\theta$ is a mean function with learnable weights. Our implementation is flexible to the choice of classes of $\mu_\theta$, so long as it is a smooth, differentiable function. Learning $\gamma$ is a non-smooth problem as it is a discrete vector. We use the Gumbel-softmax~\citep{jang:etal:2016:gumbel-softmax} to get approximate gradients for the binary mask. Both $\theta$ and $\gamma$ are fit jointly.




\subsection{Base Conditional Independence Test}
The choice of which uncalibrated conditional independence testing procedure to use is up to the user. In scenarios where there are large, unlabeled data, it may be more useful to use a model-X method. In areas where higher order moments are difficult to approximate, doubly robust methods are more likely to have higher power. We consider one method from each class.

\paragraph{Generalized Covariance Measure (GCM)~\citep{shah:peters:2020:hardness-conditional-independence}}
Fix $j$ and set $Z:=X_{-j}$. We fit estimators for the conditional mean
\[
\hat f_j(Z)\approx \mathbb{E}[X_j\mid Z],
\qquad
\hat g(Z)\approx \mathbb{E}[Y\mid Z],
\]
then form residuals 
\[
R_{X}^{(j)} \;=\; X_j-\hat f_j(Z), 
\qquad 
R_{Y}^{(j)} \;=\; Y-\hat g(Z),
\]
and elementwise products $R^{(j)}=R_{X}^{(j)}\odot R_{Y}^{(j)}$. 
Let $\bar R^{(j)}=\tfrac{1}{n}\sum_{i=1}^n R^{(j)}_i$ and 
$s_{(j)}^2=\tfrac{1}{n}\sum_{i=1}^n (R^{(j)}_i-\bar R^{(j)})^2$. 
The statistic
\[
T_j \;=\; \frac{\sqrt{n}\,\bar R^{(j)}}{s_{(j)}}
\]
is approximately $\mathcal{N}(0,1)$, yielding two-sided $p$-values $p_j=2\{1-\Phi(|T_j|)\}$. When the GCM is well-specified, it is rate doubly robust: under the null, the test statistic has the correct limiting distribution when the fitted regressions for $X_j | Z$ and $Y | Z $ converge sufficiently fast. In finite samples or with misspecification of the conditional expectations, it may inflate the error rate. 




\paragraph{Holdout Randomization Test (HRT)~\citep{tansey:etal:2021:hrt}}
Let $Z:=X_{-j}$. Under the null hypothesis, replacing $X_j$ by fresh draws from its conditional distribution given $Z$ should not worsen prediction of $Y$. The HRT uses held out prediction error as its test statistic. The HRT fits a predictor $\hat h(\cdot)$ of $Y$ from $X$ on a training split, and computes the held out loss on a holdout set ${\cal I}$,
\[
L_{\text{obs}} \;=\; \frac{1}{|{\cal I}|}\sum_{i\in{\cal I}} \ell\!\big(Y_i,\,\hat h(X_i)\big)\, .
\]
As a model-X method, the HRT estimates the conditional distribution $\hat q_j(\cdot\mid Z)$ for $X_j\mid X_{-j}$. For $b=1,\dots,B$ the HRT draws
\[
\tilde X_{j,i}^{(b)} \sim \hat q_j(\,\cdot \mid Z_i\,),\qquad i\in\{1, \ldots, n\},
\]
forms $\tilde X^{(b)}$ by replacing column $j$ on the holdout set, and computes randomized losses
\[
L^{(b)} \;=\; \frac{1}{|{\cal I}|}\sum_{i\in{\cal I}} \ell\!\big(Y_i,\,\hat h(\tilde X^{(b)}_i)\big).
\]
Compute $L_{\text{obs}}$ and $\{L^{(b)}\}_{b=1}^B$ per feature. A right–tailed $p$–value is then
\[
p_j \;=\; \frac{1+\sum_{b=1}^{B}\mathbf{1}\{\,L^{(b)}\ge L_{\text{obs}}\,\}}{B+1}.
\]
For computational efficiency, we approximate the null distribution of $L^{(b)}$ by a normal distribution using the sample mean and standard deviation of $\{L^{(b)}\}$ and compute $p$-values from this approximation.


The train-test split HRT yields valid $p$-values provided $\hat q_j$ is well specified and well estimated. 

\subsection{Calibration Metric}
Let \(\mathcal{H}_0\subseteq\{1,\dots,m\}\) denote the indices of null hypotheses, let \(m_0=|\mathcal{H}_0|\), and let \(\hat S\subseteq\{1,\dots,m\}\) denote the set of rejected hypotheses. The choice of metric to optimize against directly relates to the particular choice of error rate one is trying to control.

In the strictest case, one may wish to target the familywise error rate (FWER), i.e.,
\[
\mathrm{FWER}\;=\;\mathbb{P}\!\left(\,|\hat S\cap \mathcal{H}_0| \ge 1\,\right),
\]
which is the multiple-testing analogue of type-I error.

We consider two metrics. The first directly measures the realized type-I error. The second calibrates for the target FDR threshold.

\paragraph{Type-I.}
Define the empirical CDF of the null \(p\)-values by
\[
\widehat F_0(u)\;=\;\frac{1}{m_0}\sum_{i\in\mathcal{H}_0}\mathbf{1}\{p_i\le u\}.
\]
Fix a cutoff \(\alpha\in(0,1]\). Under perfect calibration the null \(p\)-values are uniform, so the realized type-I error at level \(\alpha\) is
\[
\widehat F_0(\alpha).
\]
We measure miscalibration by the deviation from the nominal level,
\[
\mathcal{T}(\alpha)\;=\;\widehat F_0(\alpha)-\alpha.
\]
Zero indicates exact calibration and positive values indicate inflated type-I error.

Controlling FWER is often too burdensome in large-scale testing because it requires protecting against even a single false rejection, which typically leads to very conservative thresholds and substantial loss of power. Instead, a more common target is the false discovery rate (FDR), which controls the expected fraction of false discoveries among all rejections,
\[
    \mathrm{FDR}\;=\;\mathbb{E}\!\left[ \frac{|\hat S \cap \mathcal{H}_0|}{|\hat S| \vee 1} \right].
\]
To calibrate for FDR control with BH, we use the false discovery proportion (FDP) as the miscalibration score.

\paragraph{FDP.}
At a target FDR level \(\alpha\), let \(p_{(1)}\le\cdots\le p_{(m)}\) and define
\[
t_{\mathrm{BH}} \;=\; \max\Bigl\{\tfrac{\alpha i}{m}\,:\, p_{(i)}\le \tfrac{\alpha i}{m}\Bigr\}.
\]
Reject all \(p_i\le t_{\mathrm{BH}}\). This is the Benjamini-Hochberg (BH) algorithm~\citep{benjamini:hochberg:1995:bh}.
Let \(R=\#\{i: p_i\le t_{\mathrm{BH}}\}\) and
\(V=\#\{i\in\mathcal{H}_0: p_i\le t_{\mathrm{BH}}\}\). The miscalibration score is the false discovery proportion (FDP),
\[
\mathrm{FDP} \;=\; \frac{V}{R\vee 1}.
\]
Lower values indicate better calibration with respect to BH at level \(\alpha\).


\subsection{Calibration}
Given the adversary’s worst-case metric, we construct a \emph{calibrator}, a fixed monotone map that turns raw outputs into conservative, well-calibrated ones.

Fix a calibration metric \(M\), and let \(\varphi_M(\alpha)\) denote the corresponding adversarial metric value at nominal level \(\alpha\). Define the adjusted level as
\[
\alpha_{\mathrm{cal}} \;=\; \sup\bigl\{t\in[0,1]: \varphi_M(t)\le \alpha \bigr\}.
\]
That is, we choose the largest nominal level whose realized metric under the adversary does not exceed the target \(\alpha\).

\subsection{Validity Guarantees}
The calibrator above is constructed from the worst-case adversarial value of a chosen metric over a fixed class of generators. As a result, calibrating to the worst case yields valid or conservative behavior for any other generator in the same class. We formalize this below for the FDP metric, and provide the full proof in the supplement.

\begin{theorem}
\label{thm:worstcase-validity}
Assume the true conditional law $Y\mid X$ lies in the adversary class used to compute the FDP metric function $\varphi_{\mathrm{FDP}}(\cdot)$.
Then running BH at level $\alpha_{\mathrm{cal}}$ satisfies
\[
\mathrm{FDR}\big(\mathrm{BH}(\alpha_{\mathrm{cal}})\big)\le \alpha.
\]
Moreover, $\alpha_{\mathrm{cal}}\le \alpha$, so the adjustment is conservative whenever $\varphi_{\mathrm{FDP}}(\alpha)>\alpha$.
\end{theorem}

\section{RESULTS}

We conduct a series of benchmarks to assess calibration and power for both uncalibrated and ECCIT-calibrated versions of GCM and HRT. We train and calibrate using either Type-I or FDP miscalibration metrics, and evaluate whether ECCIT calibration can restore uniform p-value distributions and control false discovery rates across different sources of miscalibration: (i) small sample sizes under a well-specified model, (ii) under-specification of exogenous noise when the true noise distribution is heavier-tailed, and (iii) model under-specification when the conditional estimators and ground truth conditionals mismatch.

In our experiments, we compare raw (uncalibrated) $p$-values to FDP and Type-I calibrated variants, evaluating power and realized FDR at $\alpha=0.2$, a standard fixed target level. On synthetic data, calibration reduces finite-sample miscalibration. The FDP calibration metric typically preserves more nominal discoveries, while the Type-I metric enforces worst-case calibration at the level of each individual test, rather than averaging error across many hypotheses. This makes the Type-I metric inherently more conservative. We also compare ECCIT-calibrated GCM and HRT against a state-of-the-art robustification method, CONTRA~\citep{sudarshan:etal:2021:contra-cit}. On semi-synthetic benchmarks, ECCIT-calibrated GCM and HRT outperform CONTRA-calibrated versions, particularly under noise or model under-specification.

\subsection{Synthetic Data Experiments}

To test calibration when conditional models are harder to estimate—e.g., under heavy tails, non-Gaussian noise, or cross-feature dependence—we evaluate several feature distributions. By default we sample independent features, i.e., \(X_{ij}\) are i.i.d.\ with zero mean:
\begin{itemize}\setlength\itemsep{2pt}
  \item \textbf{Normal:} \(X_{ij}\sim\mathcal{N}(0,1)\).
  \item \textbf{Laplace:} \(X_{ij}\sim\mathrm{Laplace}(0,1/\sqrt{2})\).
  \item \textbf{Student-t:} \(X_{ij}\sim t_{1}\).
\end{itemize}
We also test on correlated data with shared latent structures. For our \textbf{Correlated} distribution setup, we introduce a one–factor structure
\[
X \;=\; \gamma\, z\,\mathbf{1}^\top \;+\; \sqrt{1-\gamma^{2}}\,E,
\]
\[
z\sim\mathcal{N}(0,1),\; E_{ij}\sim\mathcal{N}(0,1),
\]
with \(\gamma=0.5\), yielding pairwise correlation \(\approx \gamma^{2}=0.25\).

We set $n \geq 2m$ in all runs to keep the GCM normal-equation matrices full-rank and well-conditioned, to obtain low-variance CDF/FDP maps from bootstrap resamples, and to retain enough data for testing. 

\paragraph{Responses $Y$.} To test the performance of our calibrated test, we evaluate two response models for $Y$: a sparse linear model and a simple nonlinear model.

\paragraph{Linear response.}
We use a sparse linear model with $s$ active features (chosen uniformly without replacement)
and Gaussian noise:
\[
Y \;=\; X_{S}\,\beta_{S} \;+\; \varepsilon,\quad
\beta_{S}\sim\mathcal{N}(0,I_{s}),\quad
\varepsilon\sim\mathcal{N}(0,I_{n}).
\]
The number of active features depends on $m$: we set $s=5,8,10$ active features for $m=10,25,50$, respectively.

\paragraph{Nonlinear response.}
Let \(g=\lfloor s/4\rfloor\) and split the first \(4g\) selected indices into \(g\) blocks of four, \(\{i_{b,1},i_{b,2},i_{b,3},i_{b,4}\}\). For each block we add two linear terms and one simple nonlinearity:
\[
Y \;=\;
a \sum_{b=1}^{g}\!\big(w_{b1}X_{i_{b,1}} + w_{b2}X_{i_{b,2}}\big)
\;+\;
\]
\[
b \sum_{b=1}^{g}\! u_{b}\,\tanh\!\big(c\,X_{i_{b,3}}\big)
\;+\;
\frac{a}{2}\sum_{j\in L}\! v_j X_j
\;+\; \varepsilon .
\]
Here \(w_{b1}, w_{b2}, u_b, v_j \sim \mathcal{N}(0,1)\) independently; \(a,b,c>0\) are fixed gains; \(L\) contains any remaining selected indices not used in the blocks; and \(\varepsilon \sim \mathcal{N}(0, I_n)\).
\paragraph{Model Regressors.}
We use two estimator families inside the tests and mirror the same choices in the adversary.

\textbf{Linear (Ridge).} For each feature \(j\) with \(Z:=X_{-j}\), we fit
\[
\hat f_j(z) \;=\; z^\top \hat\beta^{(x)}_j,
\qquad
\hat g(z) \;=\; z^\top \hat\beta^{(y)},
\]
where
\[
\hat\beta^{(x)}_j \;=\; (Z^\top Z+\lambda I)^{-1}Z^\top X_j,
\]
\[
\hat\beta^{(y)} \;=\; (Z^\top Z+\lambda I)^{-1}Z^\top Y,
\]
with a small ridge \(\lambda>0\) for stability.

\textbf{Nonlinear (MLP).} We replace the linear maps by a single–hidden–layer network with ReLU:
\[
h(z) \;=\; W_2\,\mathrm{ReLU}(W_1 z + b_1) + b_2,
\]
and set
\(
\hat f_j(z)=h^{(x)}_j(z),\;
\hat g(z)=h^{(y)}(z)
\)
with separate functions for predicting \(X_j\) and \(Y\) from \(Z\).

\textbf{Adversary.} The adversarial generator uses the same families to parameterize the conditional mean of \(Y\).

\paragraph{Power.} We compute power as the fraction of true features recovered after applying BH correction at level \textbf{$\alpha=0.2$}. The Type-I metric we use is also set so that $p_{max} = 0.2$. To report power only when the FDR target is met, we use \textbf{\emph{valid power}} as defined by \citet{tosh:zhang:tansey:2025:latent}:
\[
\operatorname{vp}(\alpha)=
\begin{cases}
0, & \text{if } \underline{\mathrm{FDR}}(\alpha)>\alpha,\\[4pt]
\dfrac{\#\{\text{rejected true non-nulls}\}}{\#\{\text{true non-nulls}\}}, & \text{otherwise},
\end{cases}
\]
where $\underline{\mathrm{FDR}}(\alpha)=\widehat{\mathrm{FDR}}(\alpha)-\mathrm{CI}_{\mathrm{FDR}}$ is the 95\% lower confidence bound for the average observed FDR at level $\alpha$ estimated over repeated runs. Valid power equals standard power but is set to zero whenever the FDR constraint is not satisfied. For all results on test performance, we are doing an average over 100 runs on a compute cluster, each process using 1 compute node.


\subsection{Single Hypothesis Testing}
\begin{figure}
    \centering
    \includegraphics[width=\linewidth]{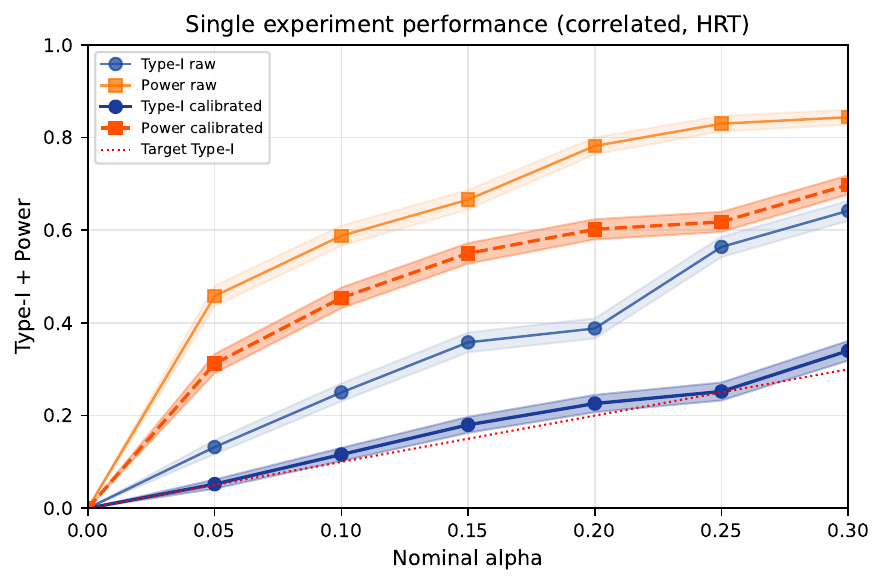}
    \caption{\textbf{Single Experiment Performance.} Realized Type-I error and power versus nominal \(\alpha\) for raw and calibrated HRT on a correlated dataset.}
    \label{fig:single_experiment}
\end{figure}

While our primary emphasis is on the multiple hypothesis setting, ECCITs also apply in the single hypothesis framework. In particular, the same calibrated mapping can be used to correct finite-sample Type-I error for a single test statistic and target level. We use the same nonlinear response construction. We first generate
\[
X = Z^\top \beta_X + \varepsilon_X,\qquad Z\in\mathbb{R}^{10},\ n=200,
\]
then set
\[
Y_{null}
= a\,(w_1 Z_{i_1}+w_2 Z_{i_2})
+ b\,u\,\tanh(Z_{i_3})
+ \varepsilon,
\]
\[
Y_{alt} = Y_0 + \tanh(X).
\]
Here we fix \(a=2\), \(b=3\), and sample coefficients $w_1, w_2, u$ with non-tiny magnitudes \(|\mathcal N(0,1)|+1\). We chose these parameters to strike a balance between model complexity and signal strength, while preserving enough signal for reliable detection. We evaluate \(\alpha\in\{0,0.05,\ldots,0.30\}\) and report realized Type-I and power before/after calibration. For the independent setting, \(Z\sim\mathcal N(0,I)\); for the correlated setting, \(Z\) gets a shared correlation structure \(Z_{ij}=\gamma U_i+\sqrt{1-\gamma^2}\,\varepsilon_{ij}\) with \(\gamma=0.5\). Figure~\ref{fig:single_experiment} shows significant improvement in Type-I error control with a tradeoff in power. Additional results are shown in the supplement.

\subsection{Sample and Feature Scaling}

\paragraph{Well-specified setting.}
For our CITs, a \emph{well specified} test is when the regression class used inside the test can correctly map the true conditionals: there exist functions \(f_j, g\) in the fitted class such that
\(f_j(x_{-j})=\mathbb{E}[X_j\mid X_{-j}=x_{-j}]\) and \(g(x_{-j})=\mathbb{E}[Y\mid X_{-j}=x_{-j}]\), and for null features \(Y\perp X_j\mid X_{-j}\). In this regime, as \(n\to\infty\) the residuals are mean-zero and the induced \(p\)-values are uniform.

Under a well-specified model, miscalibration is driven primarily by finite-sample noise and fades as \(n\) grows. Here we train the calibrator against a linear adversary and fit linear regressors in both GCM and HRT. To demonstrate this, we report calibration results across varying sample sizes and feature counts to show how our calibration results are impacted by the dimensions of our data. In this regime, both tests exhibit comparable calibrated performance.
\begin{figure*}[t!]
\centering

\begin{minipage}{0.75\textwidth}
\centering
\begin{subfigure}[t]{0.45\linewidth}
  \centering
  \includegraphics[width=\linewidth]{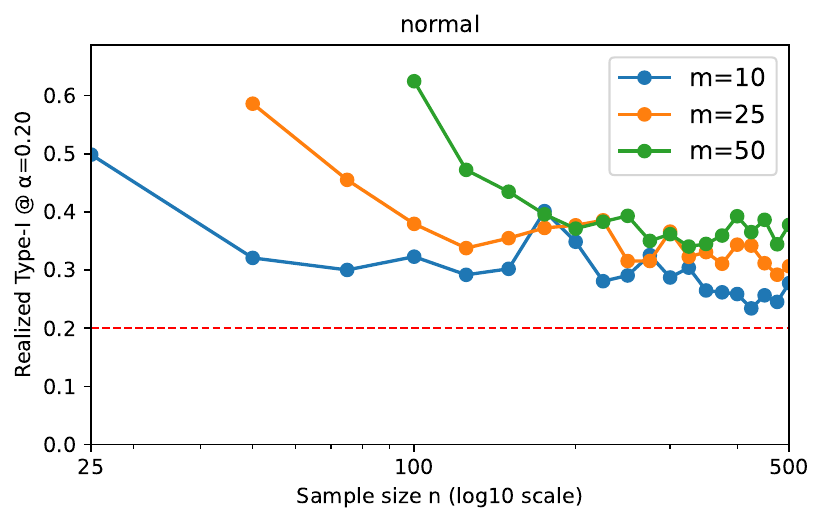}
  \caption{GCM (Type-I)}
  \label{fig:gcm_offset_normal_type1}
\end{subfigure}\hspace{1.5em}
\begin{subfigure}[t]{0.45\linewidth}
  \centering
  \includegraphics[width=\linewidth]{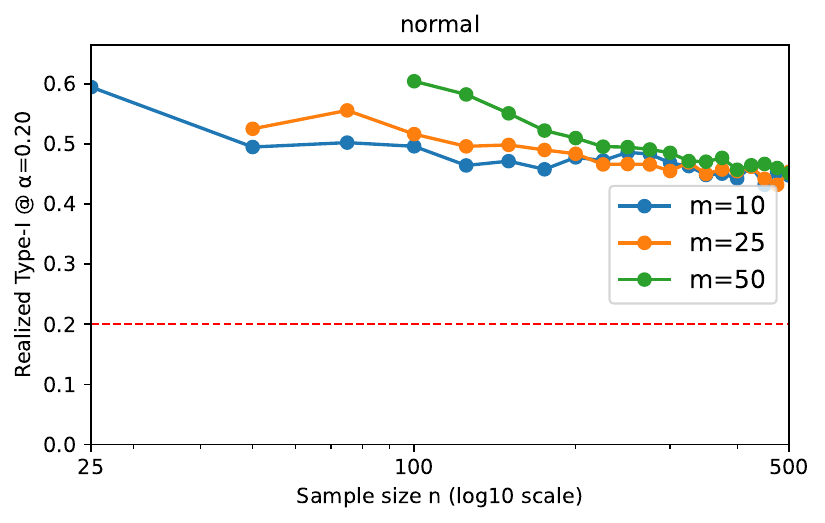}
  \caption{HRT (Type-I)}
  \label{fig:hrt_offset_normal_type1}
\end{subfigure}
\end{minipage}

\vspace{0.8em}

\begin{minipage}{0.75\textwidth}
\centering
\begin{subfigure}[t]{0.45\linewidth}
  \centering
  \includegraphics[width=\linewidth]{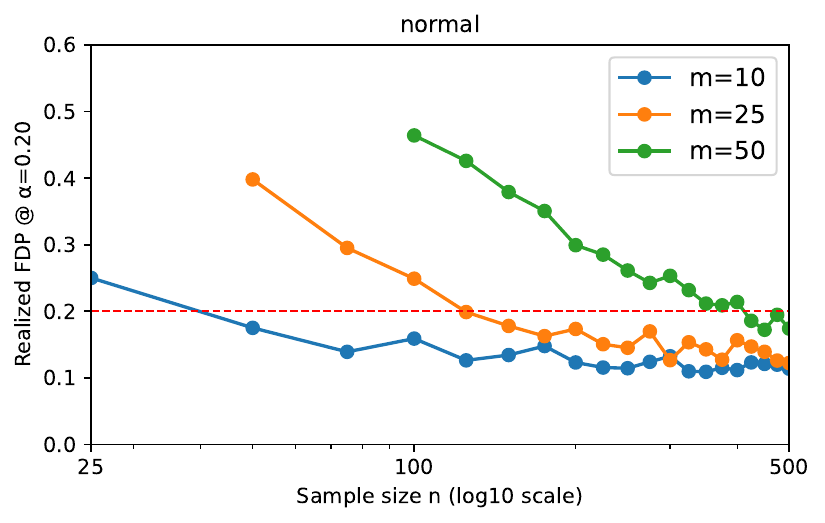}
  \caption{GCM (FDP)}
  \label{fig:gcm_offset_normal_fdp}
\end{subfigure}\hspace{1.5em}
\begin{subfigure}[t]{0.45\linewidth}
  \centering
  \includegraphics[width=\linewidth]{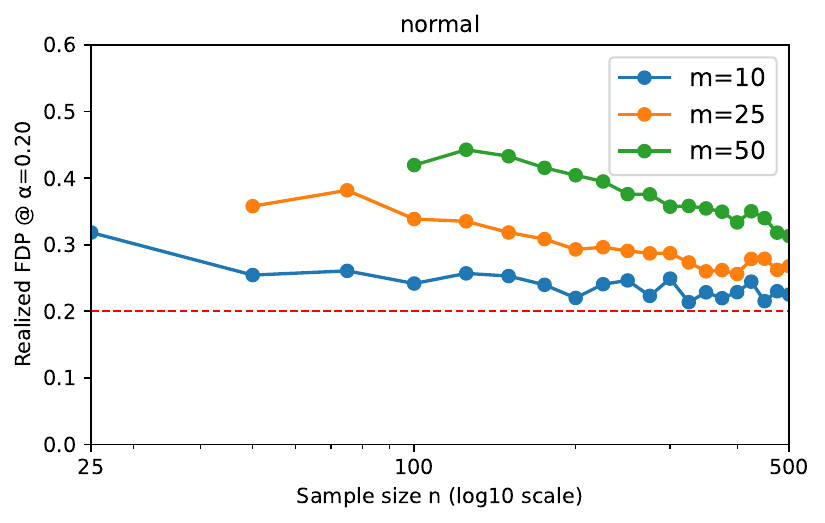}
  \caption{HRT (FDP)}
  \label{fig:hrt_offset_normal_fdp}
\end{subfigure}
\end{minipage}

\caption{\textbf{Miscalibration over sample size (log scaled) by features} on a well-specified model for both miscalibration metrics. The red dotted line indicates the selected nominal threshold of $\alpha=0.2$.}
\label{fig:offset_normal}
\end{figure*}

In Figure \ref{fig:offset_normal}, we see that even with a well specified model, both tests exhibit finite-sample deviation: for smaller $n$, the residual–product statistic is noisy and the resulting $p$-values depart from uniformity, increasingly so as $m$ grows. Our calibration here revolves around controlling for the sample noise to have a calibrated test.


\subsection{Calibration under Model Underspecification}
\paragraph{Under-specifed Setting.} The test-side regressors can be \emph{underspecified}: if the learners for $\hat f_j$ or $\hat g$ cannot capture the dependence of $Y$ on $X_{-j}$, residual structure remains and the resulting $p$-values are biased. To test this, we train the calibrator against a nonlinear adversary while the tests themselves use linear regressors, and we evaluate on data generated from a nonlinear response $Y$. For each configuration, we fit the calibrator and evaluate the performance for both the uncalibrated and calibrated tests.

\begin{figure}[t!]
\centering
\includegraphics[width=0.9\linewidth]{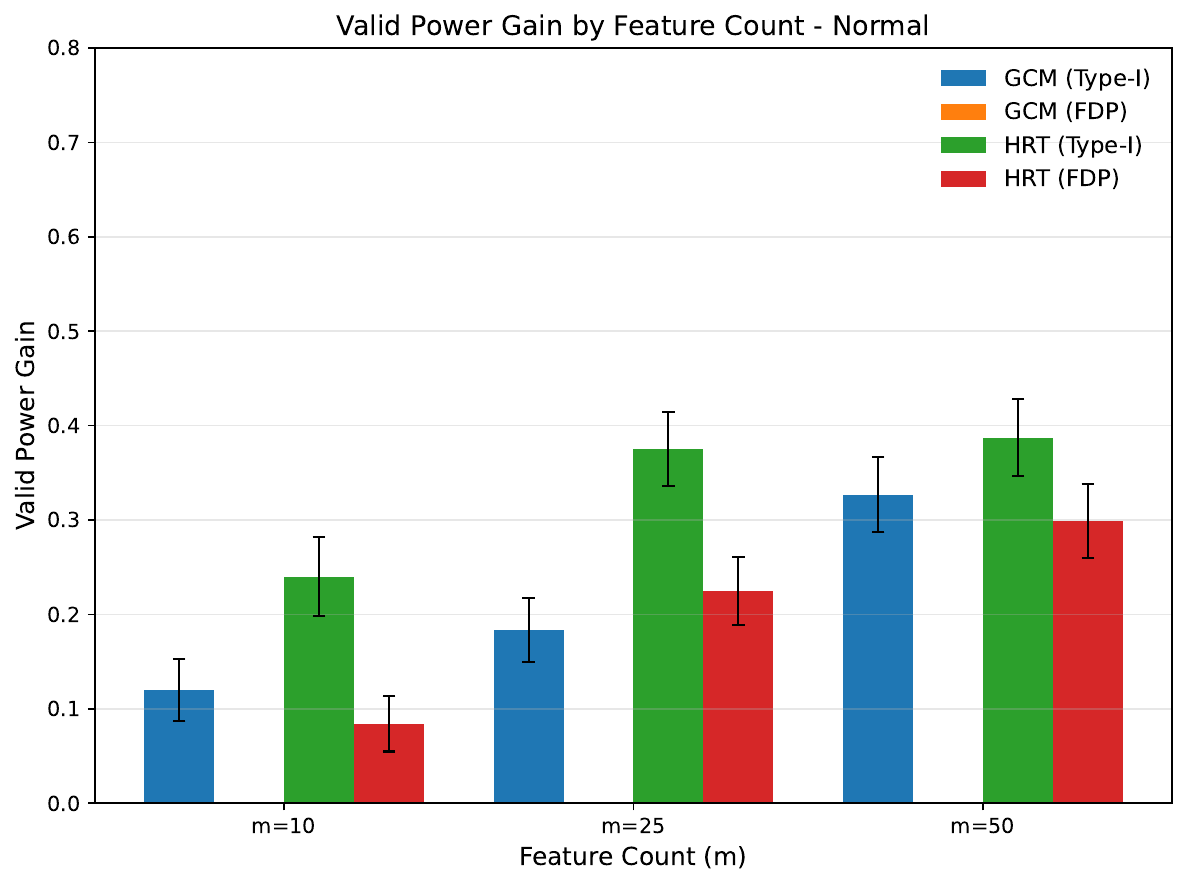}
\caption{\textbf{Valid Power Gain by Features.} Calibrated with a nonlinear adversary. Performance evaluated with $10m$ samples on a nonlinear response $Y$.}
\label{fig:features_valid_power_gain}
\end{figure}

In Figure \ref{fig:features_valid_power_gain}, the differences between the results for calibrating with our Type-I metric versus the FDP metric show us a calibration and power tradeoff. The Type-I calibrator applies a more conservative correction on $[0,\alpha]$. In the ground-truth nonlinear setting considered, this stronger correction did not substantially penalize power; however, in other response regimes the same global correction could over-adjust small $p$-values and reduce discoveries. By contrast, the FDP calibrator is targeted to BH: it selects the smallest level $\alpha_{\mathrm{cal}}$ whose realized FDP does not exceed the target, effectively bringing the threshold back to the limit. As a result, FDP calibration typically preserves more nominal power subject to the FDR constraint, whereas our Type-I metric tends to be more conservative but can provide robustness when miscalibration varies across quantiles. Again, the Type-I metric calibrates at the level of each individual test, and will drive down nominal power as a result.


\subsection{Robustness to Distributions and Noise}

\begin{figure}[t!]
\centering
\includegraphics[width=0.9\linewidth]{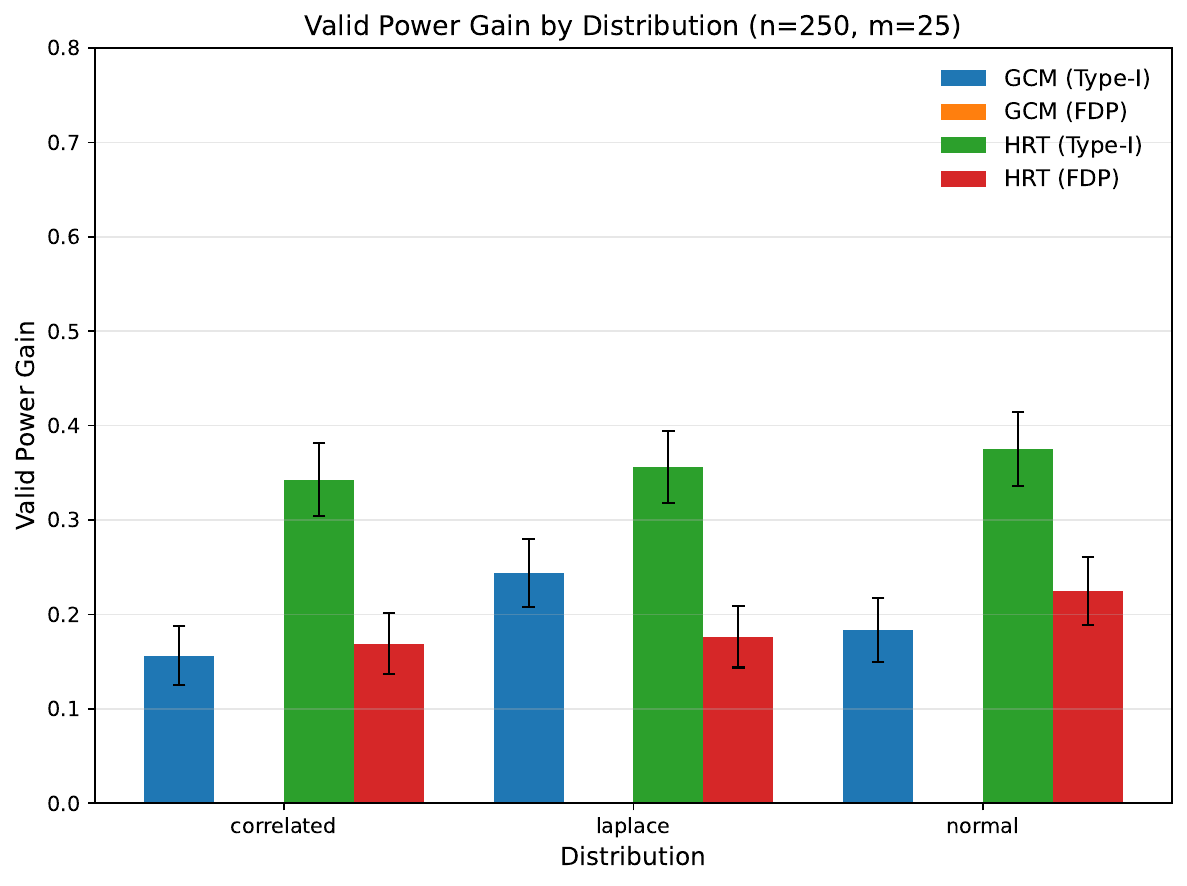}
\caption{\textbf{Valid Power Gain by Distribution.} Calibrated with a nonlinear adversary. Performance evaluated on a nonlinear response $Y$.}
\label{fig:distribution_valid_power_gain}
\end{figure}

\begin{figure*}[t!]
\centering

\begin{minipage}{0.9\textwidth}
\centering
\begin{subfigure}[t]{0.45\linewidth}
  \centering
  \includegraphics[width=\linewidth]{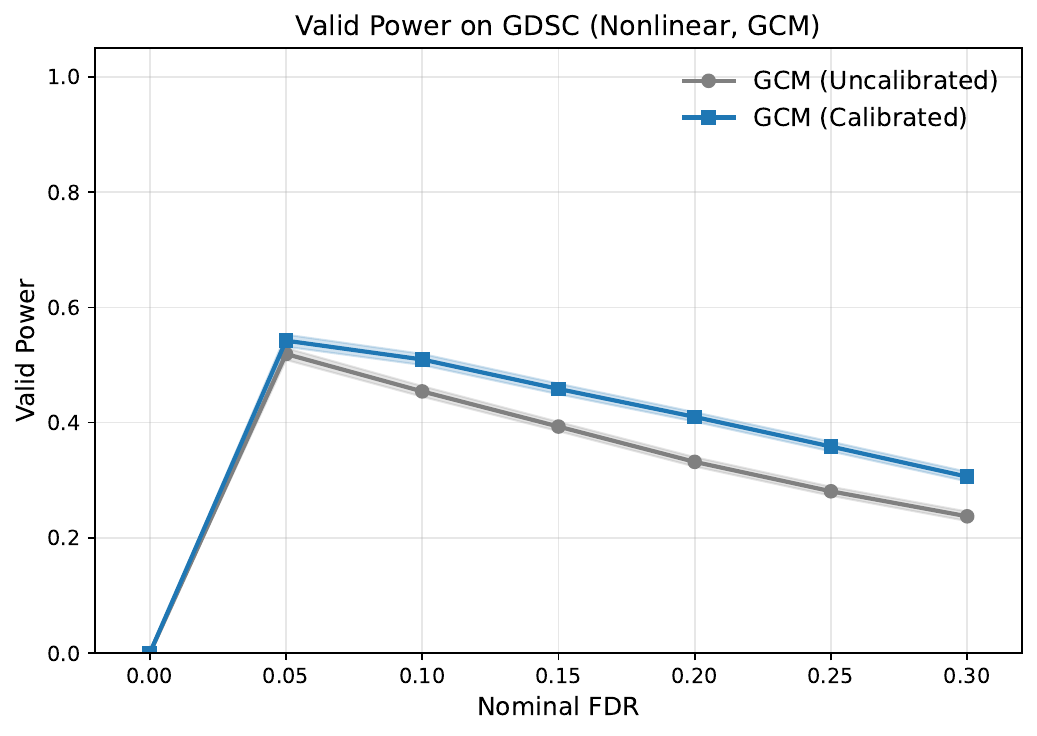}
  \label{fig:semi_gcm}
\end{subfigure}\hspace{1.5em}
\begin{subfigure}[t]{0.45\linewidth}
  \centering
  \includegraphics[width=\linewidth]{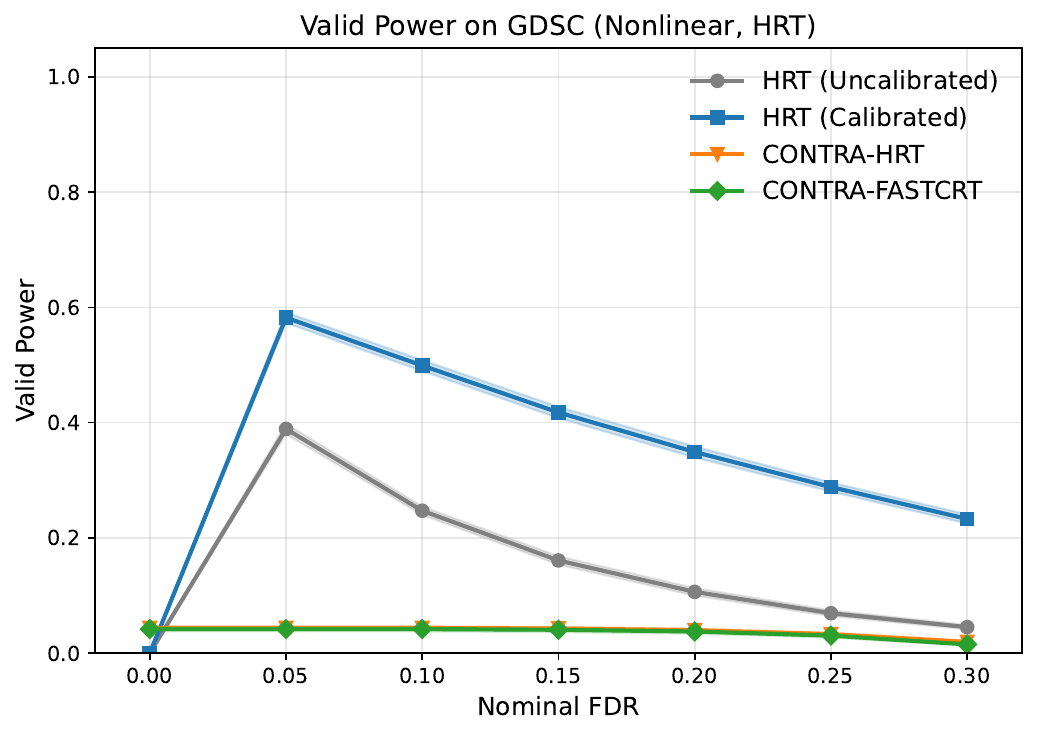}
  \label{fig:semi_hrt}
\end{subfigure}
\end{minipage}

\vspace{0.8em}

\begin{minipage}{0.9\textwidth}
\centering
\begin{subfigure}[t]{0.45\linewidth}
  \centering
  \includegraphics[width=\linewidth]{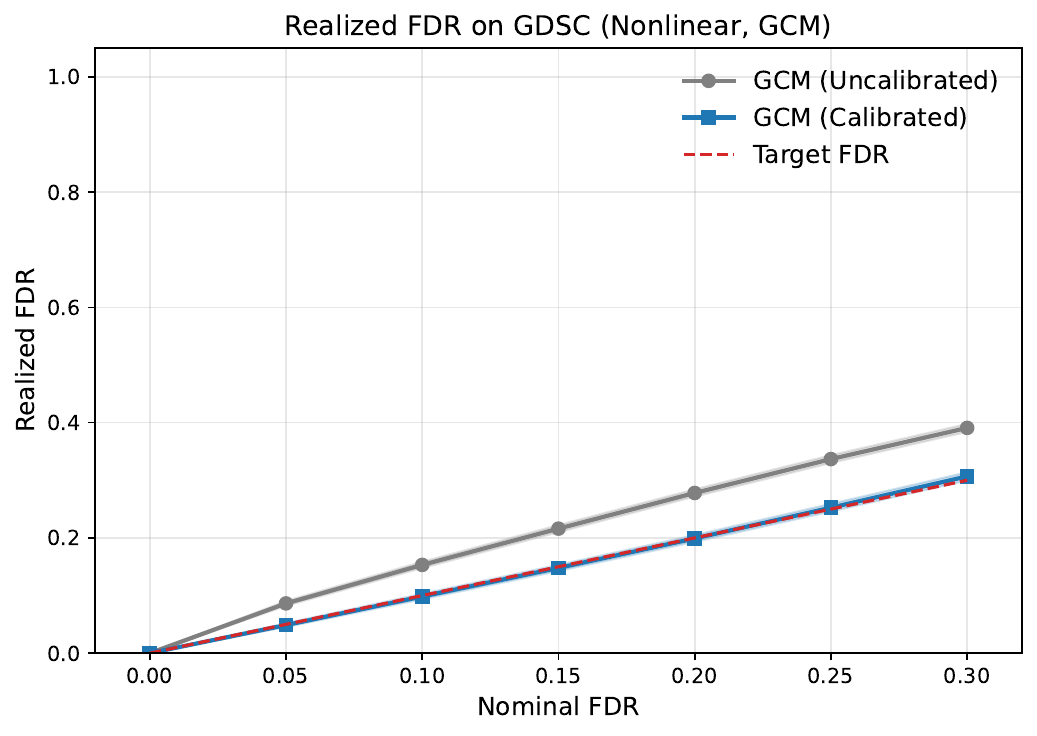}
  \label{fig:semi_gcm_fdr}
\end{subfigure}\hspace{1.5em}
\begin{subfigure}[t]{0.45\linewidth}
  \centering
  \includegraphics[width=\linewidth]{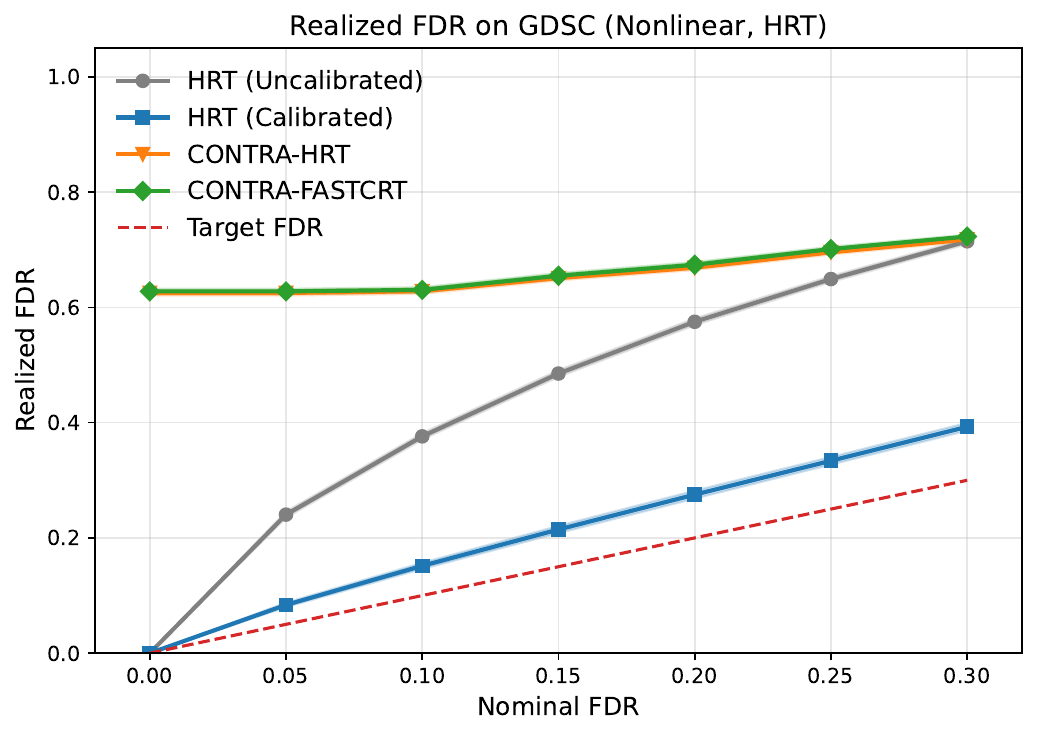}
  \label{fig:semi_hrt_fdr}
\end{subfigure}
\end{minipage}

\caption{\textbf{Valid Power and FDR Comparison on Gene Expression Data.} Calibrated with FDP metric. Nonlinear response $Y$.}
\label{fig:semi}
\end{figure*}

Latent structure and heavy–tailed noise in $X$ can degrade the conditional fits used by the tests, yielding distorted $p$-values. We set up this experiment to evaluate how well calibration restores FDR control and power when we vary the feature distribution and noise family. In this example, we keep the test side using linear regressors in GCM/HRT while training the calibrator against a nonlinear adversary and evaluating on data generated from a nonlinear response \(Y\). In Figure~\ref{fig:distribution_valid_power_gain}, we show that across different distributions and noise, we are able to improve on the tests and gain valid power through calibration.


\subsection{Comparisons on Gene Expression Data}

To illustrate a realistic use case, we apply our conditional independence tests to variable selection in gene expression analysis. This is often used due to the feature space for gene expression data. Gene features are high-dimensional and exhibit strong cross-gene correlation and latent structure, and these raw counts are often modeled with zero-inflated negative binomials. We construct a semi-synthetic benchmark from the Genomics of Drug Sensitivity in Cancer (GDSC) dataset \cite{yang:etal:2013:gdsc-nar}, treating cancer cell lines as samples (\(n\)) and genes as features (\(m\)). We normalize the expression matrix, apply a variance-stabilizing log transform, and \(z\)-score each gene. We then compare uncalibrated and calibrated GCM/HRT against CONTRA baselines under both linear and nonlinear outcome models. For each replicate, we draw a \(200\times 25\) slice, synthesize outcomes under nonlinear responses, run all methods with and without calibration, and aggregate power/FDR over 100 runs. Additional experiments and method comparisons on other datasets are reported in the supplement.

Figure \ref{fig:semi} shows that calibration restores FDR control for both GCM and HRT and increases valid power relative to baseline methods like CONTRA. We evaluate performance by sweeping the BH target level over \(\alpha \in \{0, 0.05, \ldots, 0.30\}\) and reporting FDR and power at each level. The calibrated tests track the nominal FDR and deliver consistent gains in valid power compared to our uncalibrated performance.

The CONTRA-HRT variant struggles in our high-dimensional, correlated gene-expression setting: its conditional sampler for \(X_j \mid X_{-j}\) is misspecified and the predictor is not refit under null draws, yielding miscalibrated \(p\)-values and reduced valid power. CONTRA-FASTCRT also performs poorly here because it fits only one null model per feature and reuses a fixed statistic across null resamples. Even though these tests provide very high nominal power, they have a high proportion of false discoveries that give us very low valid power.


\section{DISCUSSION}

ECCIT is a practical approach for calibrating conditional independence tests when nominal guarantees break down in practice. In our experiments, it improves calibration and increases valid power relative to existing correction methods while remaining agnostic to the base test.

A key limitation is the tradeoff between robustness and power. In ECCIT, this tradeoff is driven by the choice of adversary class, which depends on prior knowledge about the true underlying response mechanism. If the class is too simple, it may lead to insufficient correction. If it is too flexible, the worst-case calibration may become overly conservative and reduce power. A natural extension is to improve power over a set of plausible response mechanisms. We leave this direction to future work.

\bibliography{refs}

\onecolumn
\aistatstitle{Supplementary Materials}

\section{PROOF}
This is a detailed proof of Theorem 1 in the paper related to validity guarantees.

\subsection{Hypothesis testing.}

Fix a nominal alpha level $\alpha \in (0,1)$ and a multiple testing procedure $T_\alpha$ which, given data $(X,Y)$, returns a rejection set
\[
T_\alpha : (X,Y) \mapsto R_\alpha(X,Y) \subseteq \mathcal{H},
\]
where $\mathcal{H}$ is the index set of hypotheses. In the main paper, the specific choice is \(T_\alpha=\mathrm{BH}(\alpha)\), but we write the proof for a generic multiple testing procedure \(T_\alpha\).

Let $V_\alpha(X,Y)$ denote the number of false rejections among $R_\alpha(X,Y)$ under the true data generating process. We then define the false discovery proportion as
\[
\mathrm{FDP}_\alpha(X,Y)
\;:=\;
\frac{V_\alpha(X,Y)}{\max\{|R_\alpha(X,Y)|,1\}}.
\]

The corresponding false discovery rate (FDR) at level $\alpha$ is:
\[
\mathrm{FDR}_\alpha
\;:=\;
\mathbb{E}\big[\mathrm{FDP}_\alpha(X,Y)\big].
\]

\subsection{FDP loss.}

We first define the pointwise FDP loss at level $\alpha$:
\[
\ell_\alpha(Y)
\;:=\;
\big(\mathrm{FDP}_\alpha(X,Y) - \alpha\big)_+
\;=\;
\max\{\mathrm{FDP}_\alpha(X,Y) - \alpha,\, 0\},
\]
so that $\ell_\alpha(Y) = 0$ whenever $\mathrm{FDP}_\alpha(X,Y) \le \alpha$, and $\ell_\alpha(Y)$ measures the amount by which we exceed level $\alpha$ otherwise.

\subsection{Adversarial model and miscalibration.}

In our calibration procedure, we consider a class of adversary functions $\mathcal{F}$ (for example, linear maps $f: \mathbb{R}^p \to \mathbb{R}$). For each $f \in \mathcal{F}$, the adversary specifies a conditional distribution $P_f(Y \mid X)$ (e.g., $Y = f(X) + \varepsilon$ with a fixed noise model). We write $Y^f \sim P_f(\cdot \mid X)$ for a random outcome generated under this adversarial model.

We then define the miscalibration of $f$ at nominal level $\alpha$ as
\[
L_\alpha(f)
\;:=\;
\mathbb{E}\big[\ell_\alpha(Y^f)\big]
\;=\;
\mathbb{E}\big[(\mathrm{FDP}_\alpha(X,Y^f) - \alpha)_+\big].
\]

The ideal adversarial objective over this class is the worst case miscalibration
\[
M_{\mathcal{F}}
\;:=\;
\sup_{f \in \mathcal{F}} L_\alpha(f),
\]
the largest expected excess FDP among all generators $f \in \mathcal{F}$.

In practice, the adversary training only approximately maximizes $L_\alpha(f)$ over $f \in \mathcal{F}$. If $\hat f$ denotes the final adversary we obtain, we assume an optimization error $\epsilon_{\mathrm{opt}}$ such that
\[
0
\;\le\;
M_{\mathcal{F}} - L_\alpha(\hat f)
\;\le\;
\epsilon_{\mathrm{opt}},
\]
so that
\[
L_\alpha(\hat f)
\;\ge\;
M_{\mathcal{F}} - \epsilon_{\mathrm{opt}},
\]
\begin{equation}
M_{\mathcal{F}}
\;\le\;
L_\alpha(\hat f) + \epsilon_{\mathrm{opt}}.
\label{eq:adversary-dominance}
\end{equation}

\subsection{Bounds for a fixed $f$.}

For any fixed $f \in \mathcal{F}$ and any $\delta > 0$,
\[
\big\{\mathrm{FDP}_\alpha(X,Y^f) \ge \alpha + \delta\big\}
\;\subseteq\;
\big\{\big(\mathrm{FDP}_\alpha(X,Y^f) - \alpha\big)_+ \ge \delta\big\},
\]
since the event on the left implies that the excess $(\mathrm{FDP}_\alpha - \alpha)_+$ is at least $\delta$. Therefore, by Markov's inequality,
\begin{align*}
\mathbb{P}_f\big(\mathrm{FDP}_\alpha(X,Y^f) \ge \alpha + \delta\big)
&\le
\frac{\mathbb{E}\big[(\mathrm{FDP}_\alpha(X,Y^f) - \alpha)_+\big]}{\delta}
= \frac{L_\alpha(f)}{\delta}.
\end{align*}
This gives a tail bound for the FDP under the adversarial model $P_f$.
\newline

Similarly, the expected FDP under $P_f$ satisfies
\begin{align*}
\mathbb{E}_f\big[\mathrm{FDP}_\alpha(X,Y^f)\big]
&=
\alpha
+ \mathbb{E}_f\big[\mathrm{FDP}_\alpha(X,Y^f) - \alpha\big] \\
&\le
\alpha
+ \mathbb{E}_f\big[(\mathrm{FDP}_\alpha(X,Y^f) - \alpha)_+\big] \\
&=
\alpha + L_\alpha(f),
\end{align*}
so $L_\alpha(f)$ also controls the FDR for the model $P_f$ as an excess over the nominal level $\alpha$.

\subsection{Bounds when the ground truth lies in $\mathcal{F}$.}

Suppose now that the true conditional distribution $P_{\mathrm{true}}(Y \mid X)$ belongs to our adversary class, in the sense that there exists some $f_{\mathrm{true}} \in \mathcal{F}$ such that
\[
P_{\mathrm{true}}(\,\cdot \mid X)
\;=\;
P_{f_{\mathrm{true}}}(\,\cdot \mid X).
\]
Let $Y_{\mathrm{true}} \sim P_{\mathrm{true}}(\cdot \mid X)$ denote the true response. By construction, $Y_{\mathrm{true}}$ and $Y^{f_{\mathrm{true}}}$ have the same conditional distribution given $X$, so any bound that holds under $P_{f_{\mathrm{true}}}$ also holds under $P_{\mathrm{true}}$.

By definition of supremum, and combining \eqref{eq:adversary-dominance} with $f = f_{\mathrm{true}}$,
\[
L_\alpha(f_{\mathrm{true}})
\;\le\;
M_{\mathcal{F}},
\]
\begin{equation}
L_\alpha(f_{\mathrm{true}})
\;\le\;
L_\alpha(\hat f) + \epsilon_{\mathrm{opt}}.
\label{eq:true-Lfstar}
\end{equation}
Applying the tail bound above with $f = f_{\mathrm{true}}$ gives, for any $\delta > 0$,
\begin{align*}
\mathbb{P}_{\mathrm{true}}\big(\mathrm{FDP}_\alpha(X,Y_{\mathrm{true}}) \ge \alpha + \delta\big)
&=
\mathbb{P}_{f_{\mathrm{true}}}\big(\mathrm{FDP}_\alpha(X,Y^{f_{\mathrm{true}}}) \ge \alpha + \delta\big) \\
&\le
\frac{L_\alpha(f_{\mathrm{true}})}{\delta} \\
&\le
\frac{L_\alpha(\hat f) + \epsilon_{\mathrm{opt}}}{\delta}.
\end{align*}
Similarly, for the expected FDP under the true model,
\begin{align*}
\mathbb{E}_{\mathrm{true}}\big[\mathrm{FDP}_\alpha(X,Y_{\mathrm{true}})\big]
&=
\mathbb{E}_{f_{\mathrm{true}}}\big[\mathrm{FDP}_\alpha(X,Y^{f_{\mathrm{true}}})\big] \\
&\le
\alpha + L_\alpha(f_{\mathrm{true}}) \\
&\le
\alpha + L_\alpha(\hat f) + \epsilon_{\mathrm{opt}}.
\end{align*}

Thus, under the assumption that the true conditional distribution $Y \mid X$ lies in our adversary class $\mathcal{F}$, the optimized adversarial objective $L_\alpha(\hat f)$ provides an upper bound, up to the optimization error $\epsilon_{\mathrm{opt}}$, on both the probability of large FDP ranges and on the FDR of our procedure.

\subsection{Calibration.}

For each nominal level $\alpha \in (0,1)$ we have the bound
\[
B(\alpha)
\;:=\;
\alpha + L_\alpha(\hat f) + \epsilon_{\mathrm{opt}}.
\]
To match the notation in the main text, we may equivalently write
\[
\varphi_{\mathrm{FDP}}(\alpha) \;:=\; B(\alpha).
\]
Since $L_\alpha(\hat f) \ge 0$ and $\epsilon_{\mathrm{opt}} \ge 0$, the calibration curve $B(\alpha)$ is pointwise lower bounded by the identity map,
\[
B(\alpha) \;\ge\; \alpha \quad\text{for all }\alpha \in (0,1).
\]

Given a desired target FDR level $q \in (0,1)$, define the calibrated nominal level as the largest nominal level whose worst-case FDR bound does not exceed $q$:
\[
\alpha_{\mathrm{cal}}(q)
\;:=\;
\sup\big\{t \in [0,1] : B(t) \le q\big\}.
\]
By construction,
\[
B(\alpha_{\mathrm{cal}}(q)) \;\le\; q,
\]
and therefore
\[
\mathbb{E}_{\mathrm{true}}\big[\mathrm{FDP}_{\alpha_{\mathrm{cal}}(q)}(X,Y_{\mathrm{true}})\big]
\;\le\;
B(\alpha_{\mathrm{cal}}(q))
\;\le\;
q.
\]
The calibrated nominal level is always less than or equal to the target FDR level. In particular, this guarantees that the calibration step can only reduce, or leave unchanged, the effective FDR of the procedure under the true data generating process.

\clearpage

\section{SINGLE EXPERIMENTS}
\begin{figure}
\begin{subfigure}[t]{0.45\textwidth}
    \centering
    \includegraphics[width=\linewidth]{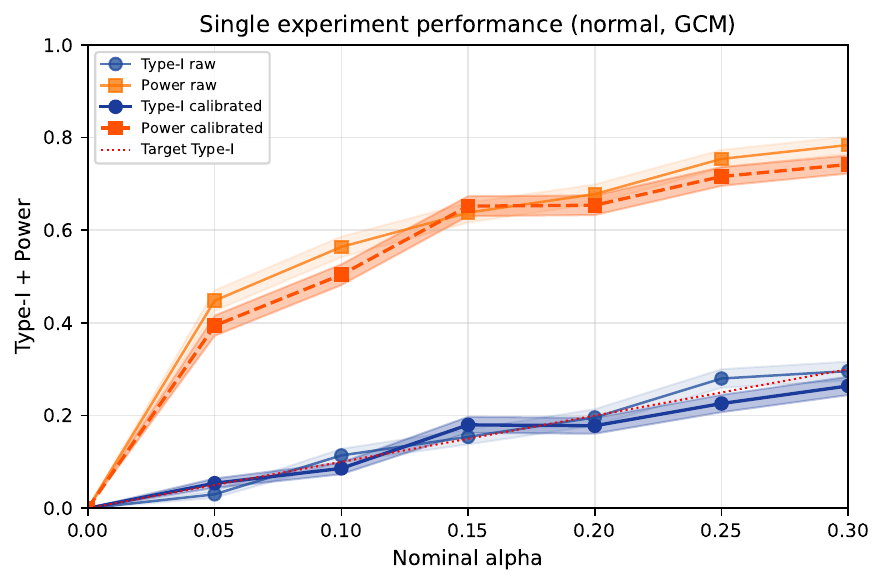}
\end{subfigure}\hfill
\begin{subfigure}[t]{0.45\textwidth}
    \centering
    \includegraphics[width=\linewidth]{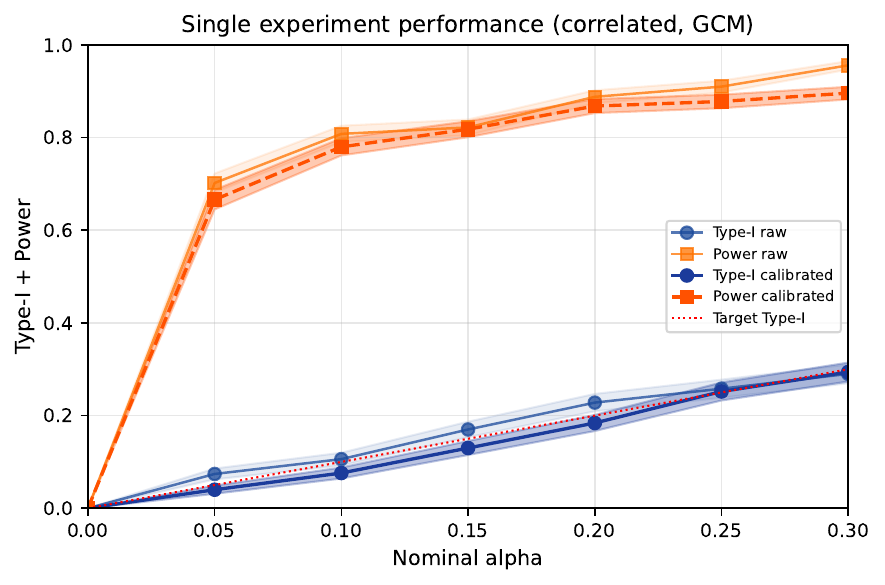}
\end{subfigure}
\medskip
\begin{subfigure}[t]{0.45\textwidth}
    \centering
    \includegraphics[width=\linewidth]{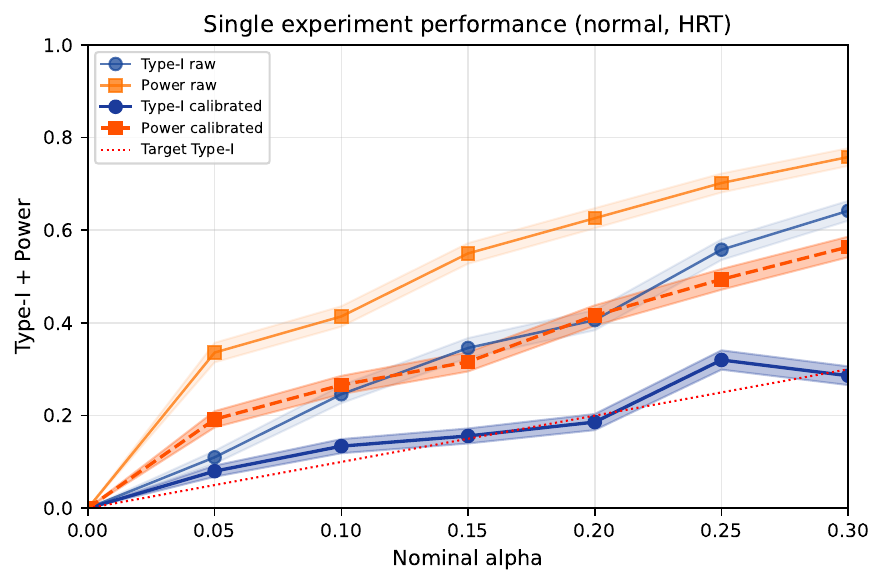}
\end{subfigure}\hfill
\begin{subfigure}[t]{0.45\textwidth}
    \centering
    \includegraphics[width=\linewidth]{figures/single_experiments/single_correlated_hrt.pdf}
\end{subfigure}
    \caption{\textbf{Single Experiment Performance.} Realized Type-I error and power versus nominal \(\alpha\) for raw and calibrated procedures in independent and correlated settings.}
    \label{fig:single_experiments}
\end{figure}

Figure~\ref{fig:single_experiments} shows additional results for the independent and correlated Gaussian settings using GCM and HRT.

\clearpage

\section{EXPANDED RESULTS}
The results below expand on the experiments shown in the main paper with more examples, in particular for different sample sizes and features, as well as a comparison across distributions and adversary choice.

\begin{figure}[t]
\begin{subfigure}[t]{0.49\textwidth}
    \centering
    \includegraphics[width=\linewidth]{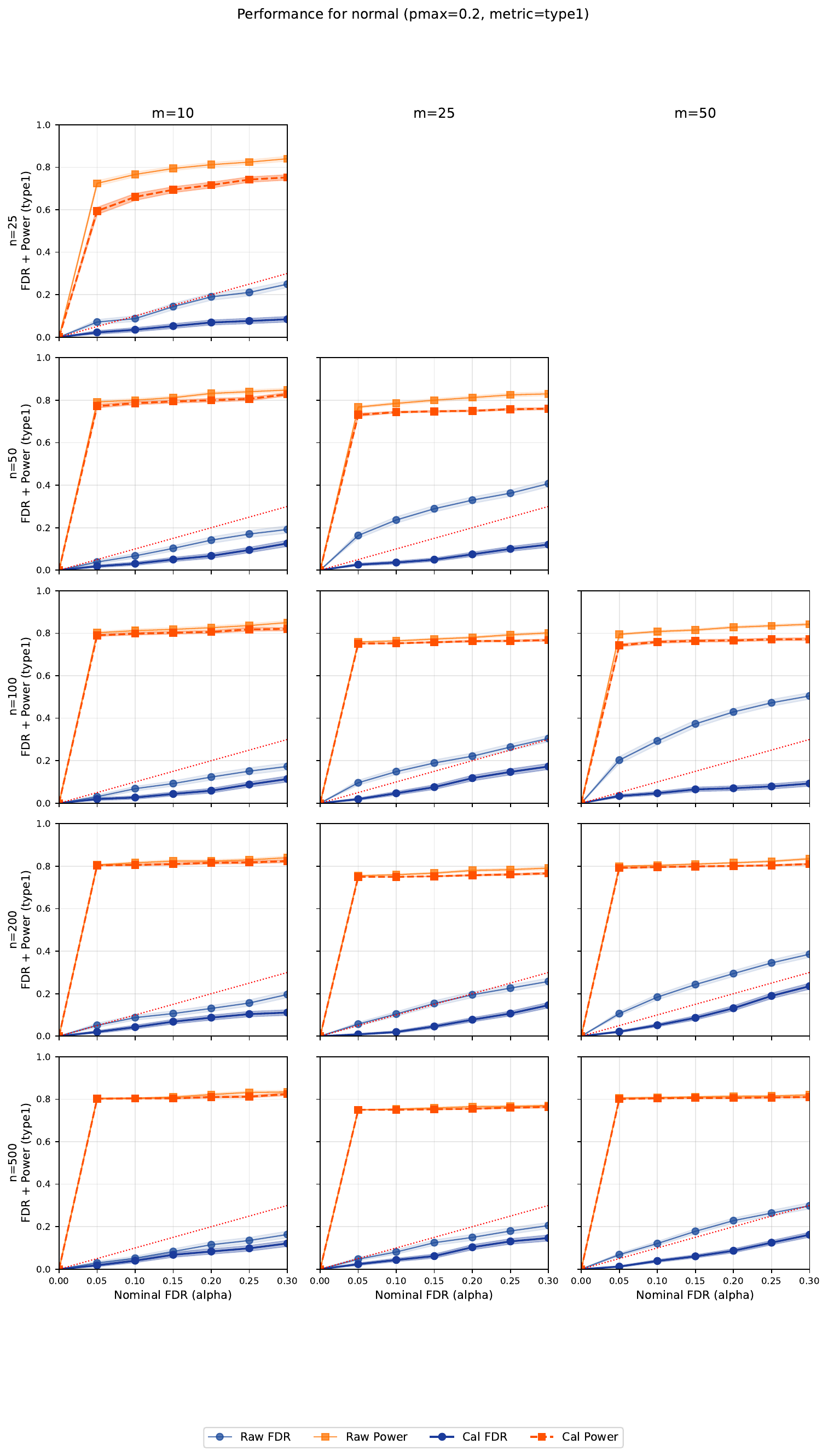}
\end{subfigure}
\begin{subfigure}[t]{0.49\textwidth}
    \centering
    \includegraphics[width=\linewidth]{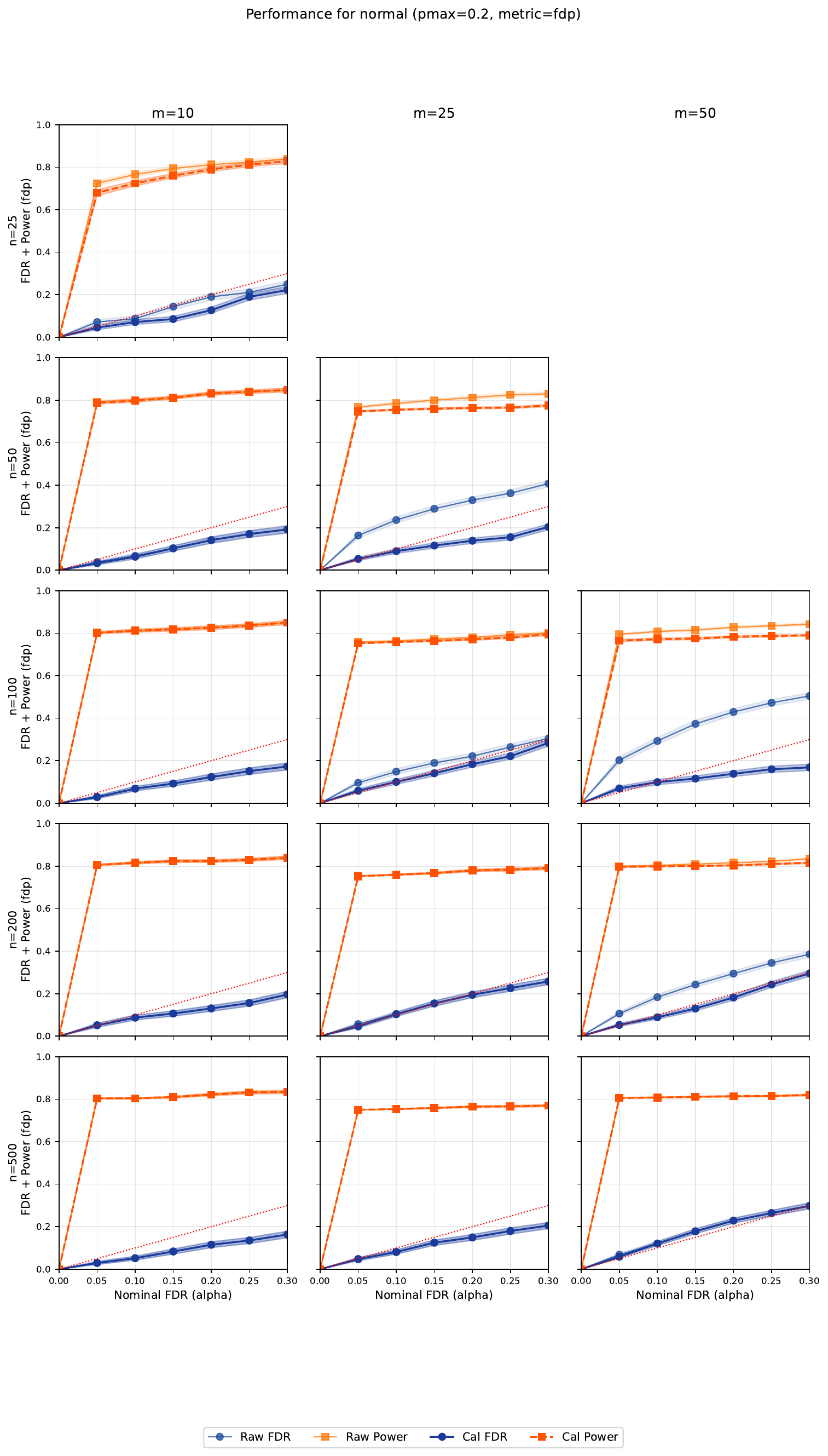}
\end{subfigure}
    \caption{\textbf{GCM Performance.} Calibrated with a nonlinear adversary with both metrics. Performance evaluated on a nonlinear ground truth response. Results show the change in power and FDR before and after calibration across different numbers of samples and features.}
    \label{fig:gcm_performance}
\end{figure}

\begin{figure}
\begin{subfigure}[t]{0.49\textwidth}
    \centering
    \includegraphics[width=\linewidth]{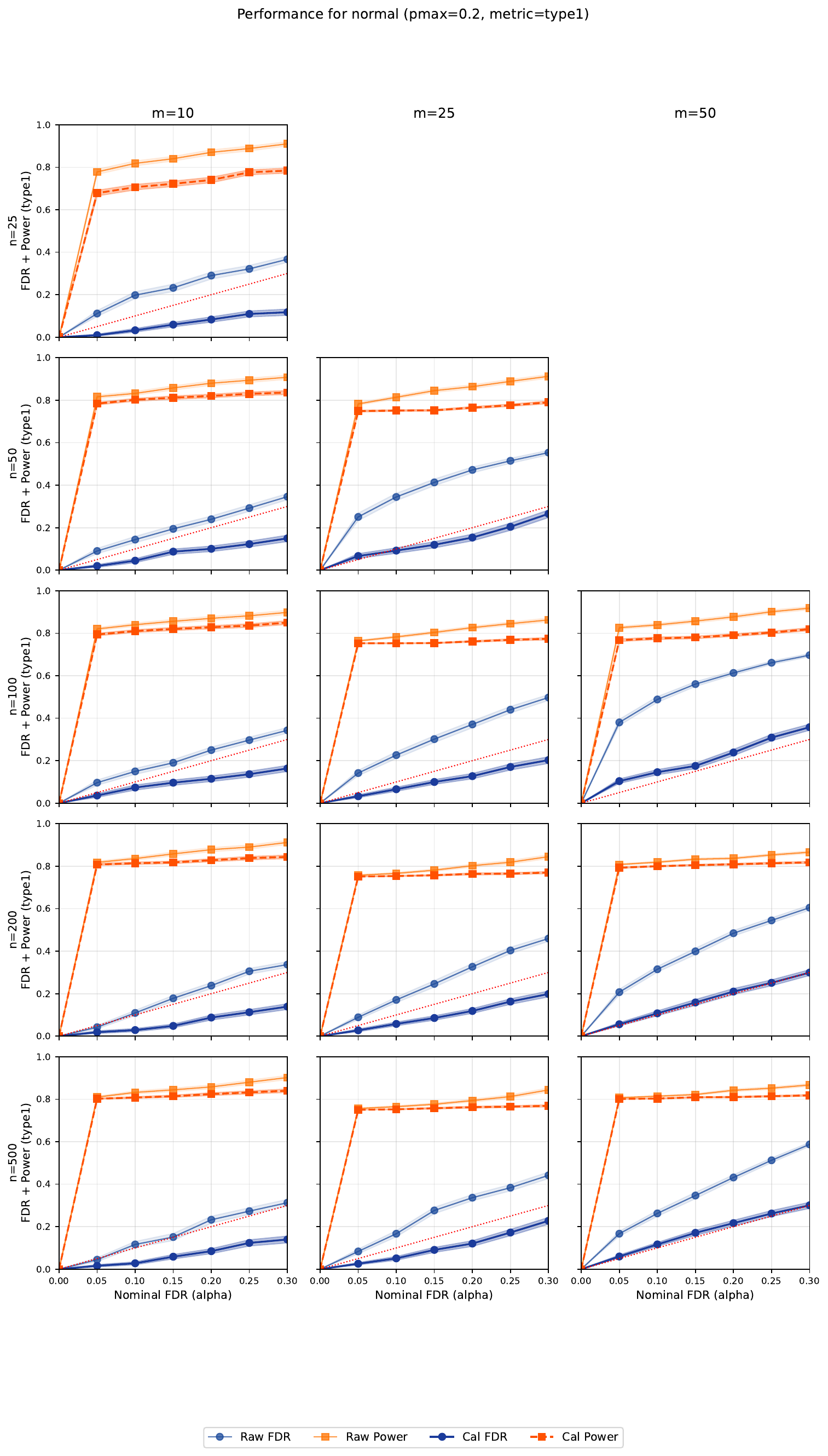}
\end{subfigure}
\begin{subfigure}[t]{0.49\textwidth}
    \centering
    \includegraphics[width=\linewidth]{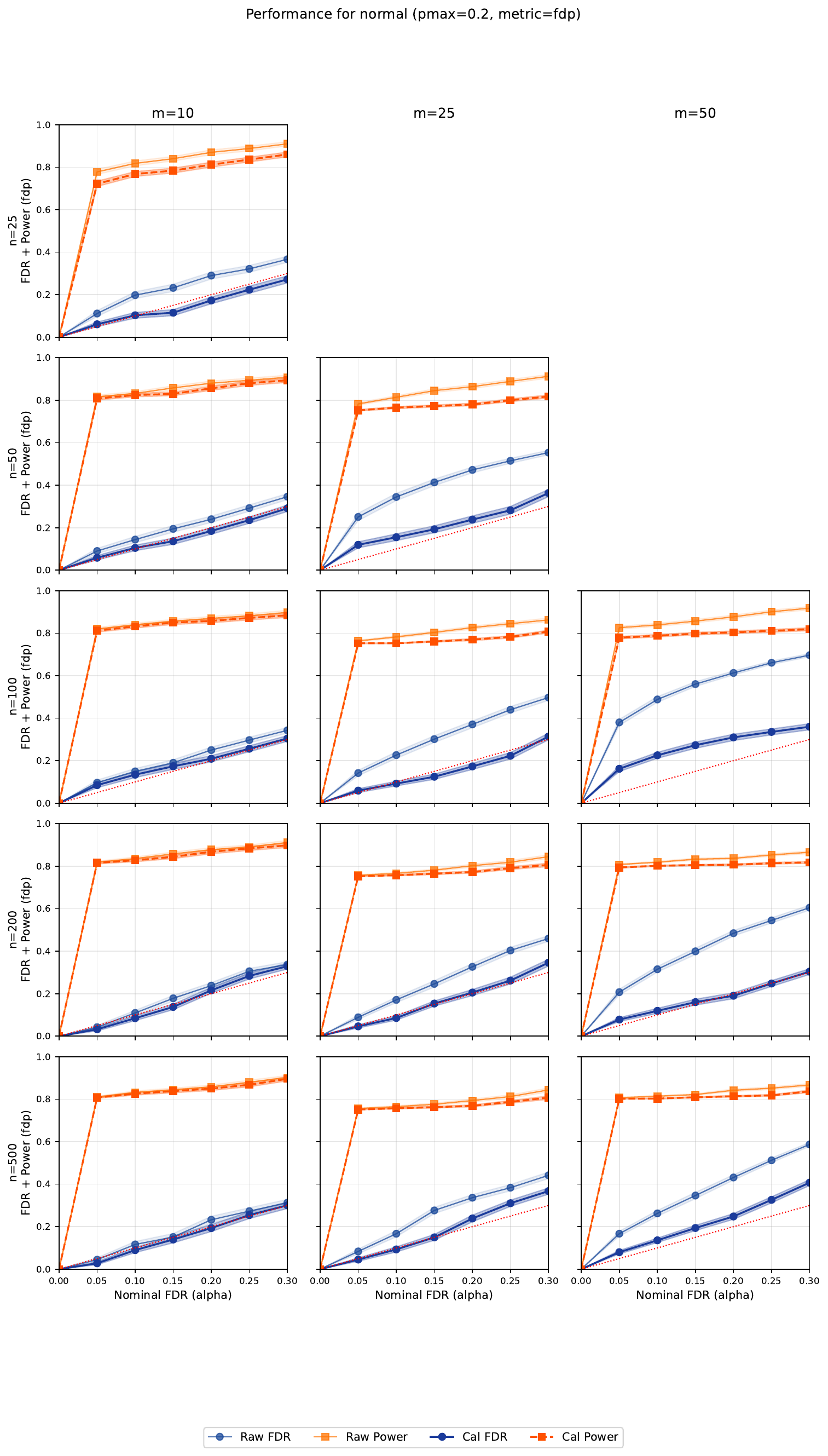}
\end{subfigure}
    \caption{\textbf{HRT Performance.} Calibrated with a nonlinear adversary with both metrics. Performance evaluated on a nonlinear ground truth response. Results show the change in power and FDR before and after calibration across different numbers of samples and features.}
    \label{fig:hrt_performance}
\end{figure}

\begin{figure}[t!]
\centering
\begin{subfigure}[t]{0.33\textwidth}
  \centering
  \includegraphics[width=\linewidth]{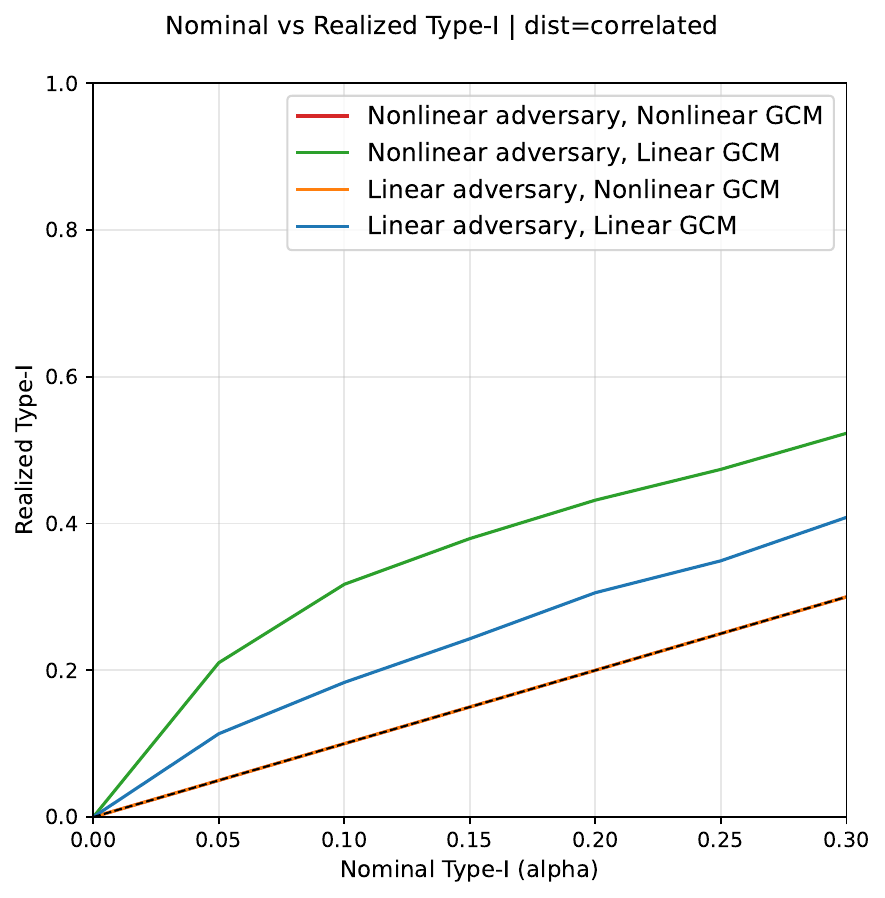}
  \caption{Correlated (Type-I)}
  \label{fig:gcm_type1_corr}
\end{subfigure}\hfill
\begin{subfigure}[t]{0.33\textwidth}
  \centering
  \includegraphics[width=\linewidth]{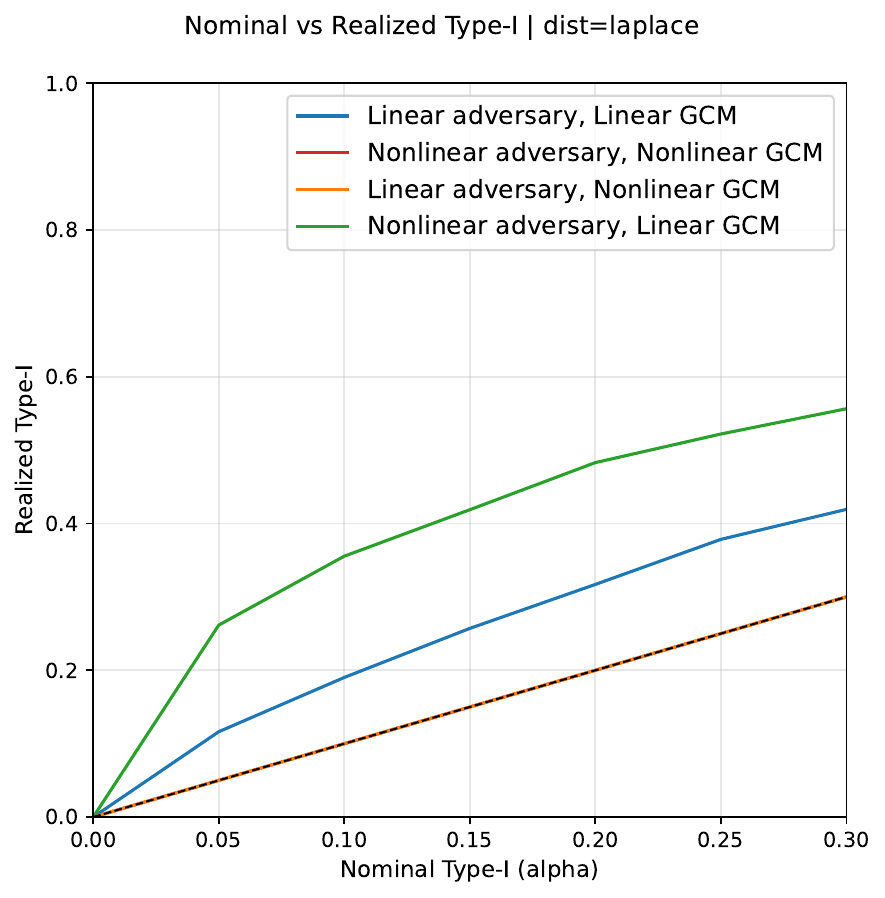}
  \caption{Laplace (Type-I)}
  \label{fig:gcm_type1_lap}
\end{subfigure}\hfill
\begin{subfigure}[t]{0.33\textwidth}
  \centering
  \includegraphics[width=\linewidth]{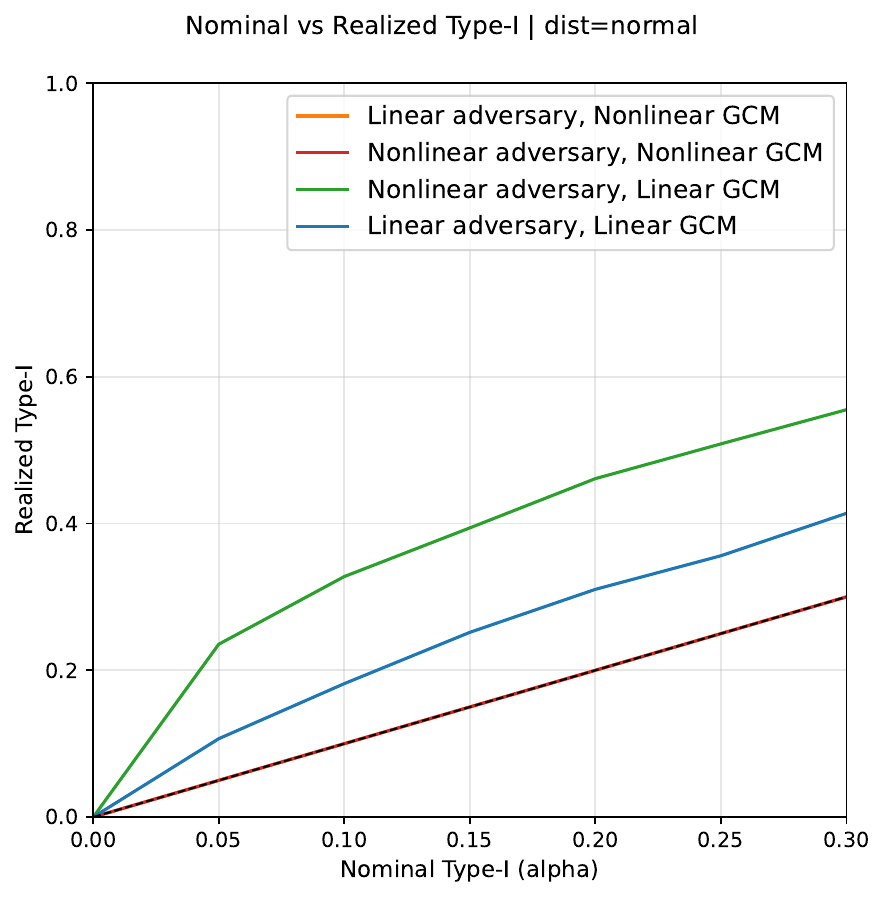}
  \caption{Normal (Type-I)}
  \label{fig:gcm_type1_norm}
\end{subfigure}\hfill

\vspace{0.6em}

\begin{subfigure}[t]{0.33\textwidth}
  \centering
  \includegraphics[width=\linewidth]{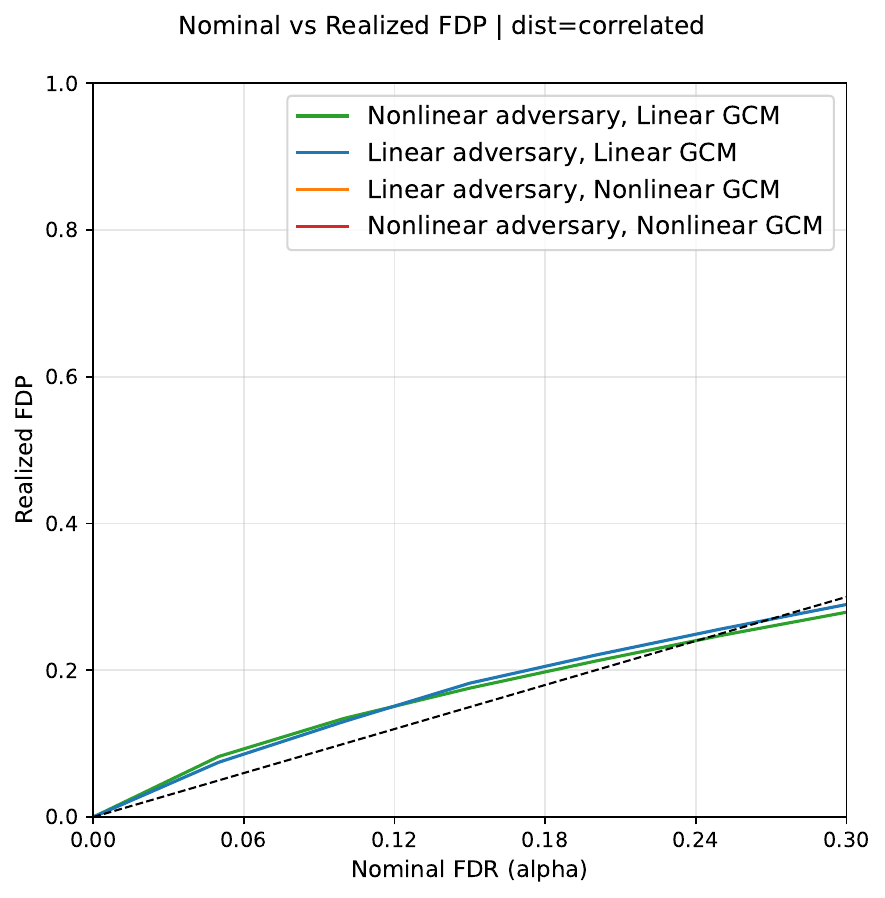}
  \caption{Correlated (FDP)}
  \label{fig:gcm_fdp_corr}
\end{subfigure}\hfill
\begin{subfigure}[t]{0.33\textwidth}
  \centering
  \includegraphics[width=\linewidth]{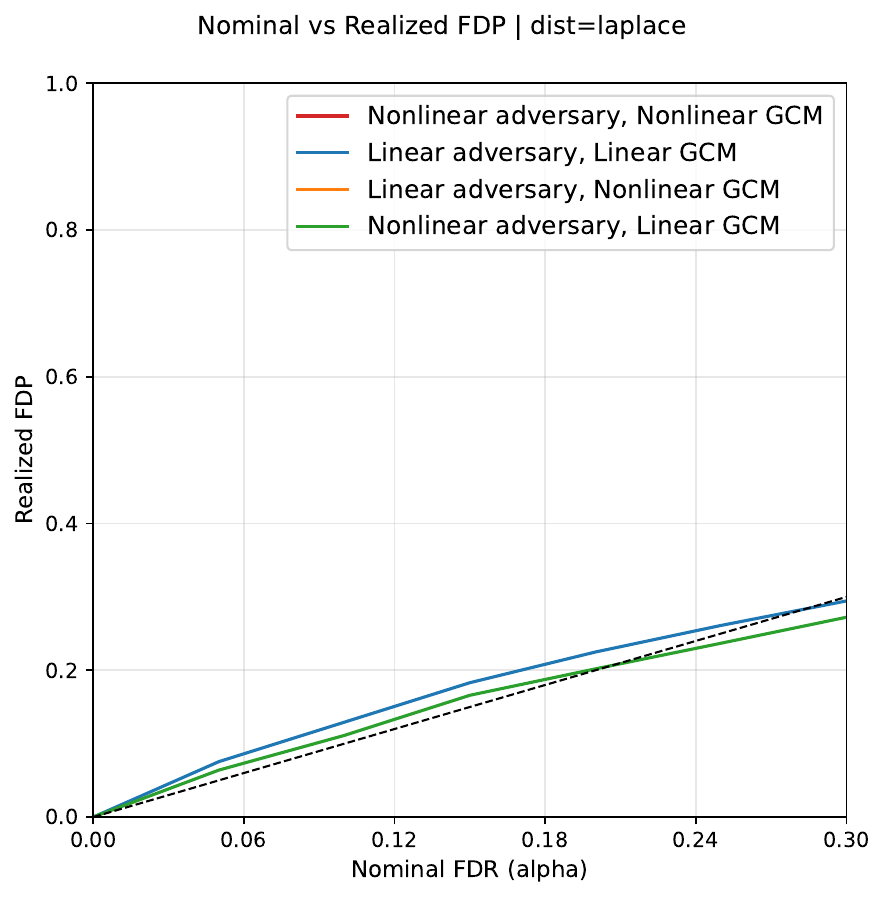}
  \caption{Laplace (FDP)}
  \label{fig:gcm_fdp_lap}
\end{subfigure}\hfill
\begin{subfigure}[t]{0.33\textwidth}
  \centering
  \includegraphics[width=\linewidth]{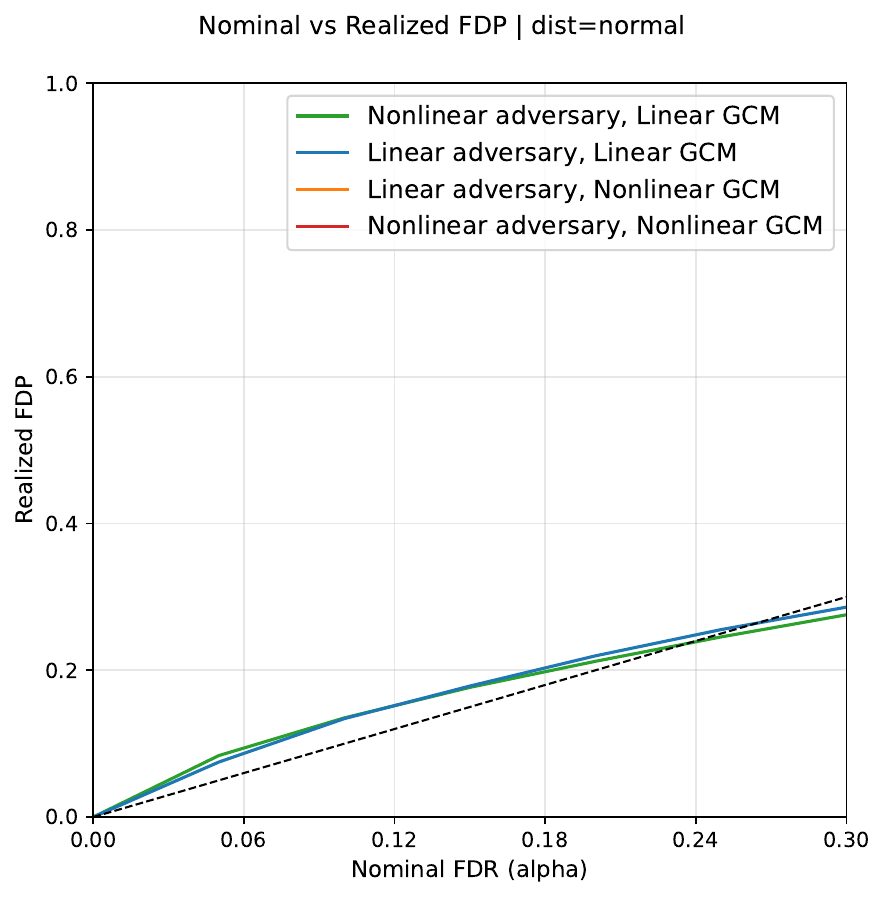}
  \caption{Normal (FDP)}
  \label{fig:gcm_fdp_norm}
\end{subfigure}\hfill

\caption{\textbf{GCM calibration across distributions.} Comparisons between different miscalibration combinations on the adversary and the test. Type-I metric (top) and FDP metric (bottom).}
\label{fig:gcm_cdf_fdp_grid}
\end{figure}

\begin{figure}[t!]
\centering
\begin{subfigure}[t]{0.33\textwidth}
  \centering
  \includegraphics[width=\linewidth]{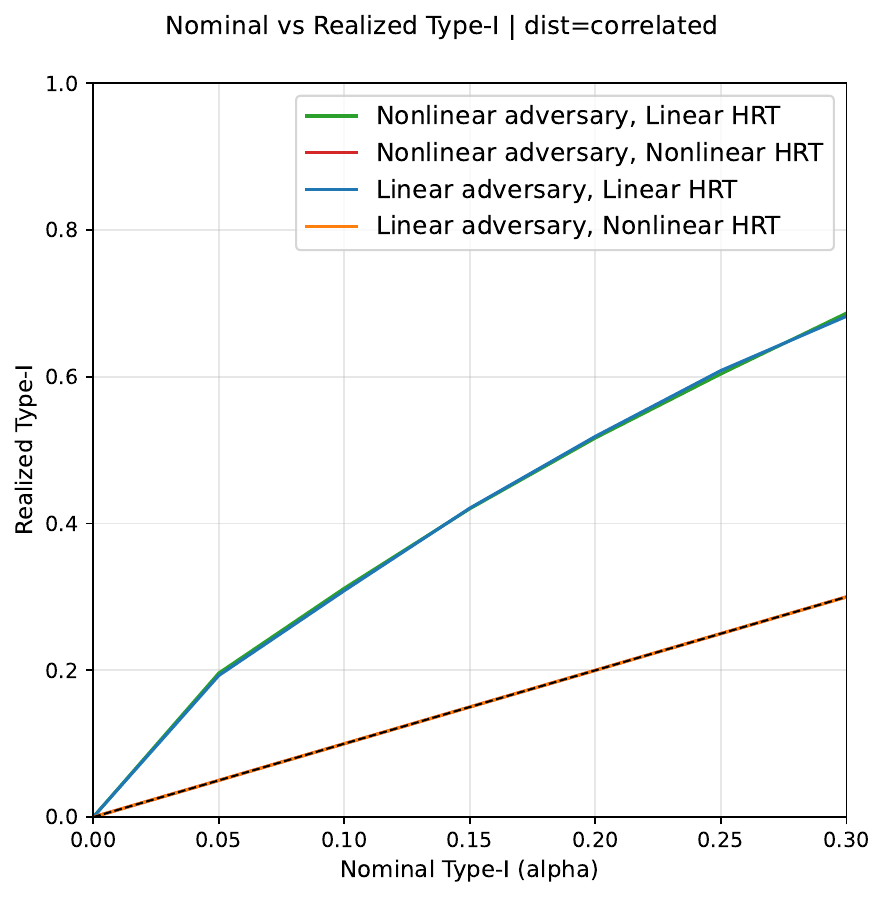}
  \caption{Correlated (Type-I)}
  \label{fig:hrt_type1_corr}
\end{subfigure}\hfill
\begin{subfigure}[t]{0.33\textwidth}
  \centering
  \includegraphics[width=\linewidth]{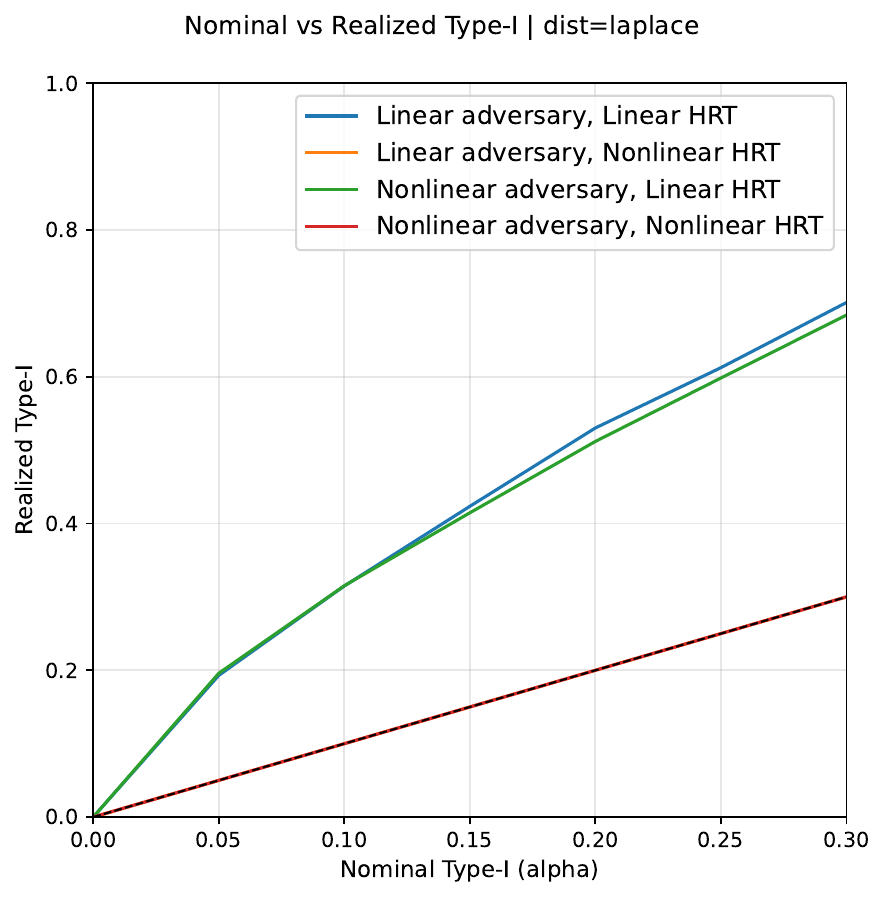}
  \caption{Laplace (Type-I)}
  \label{fig:hrt_type1_lap}
\end{subfigure}\hfill
\begin{subfigure}[t]{0.33\textwidth}
  \centering
  \includegraphics[width=\linewidth]{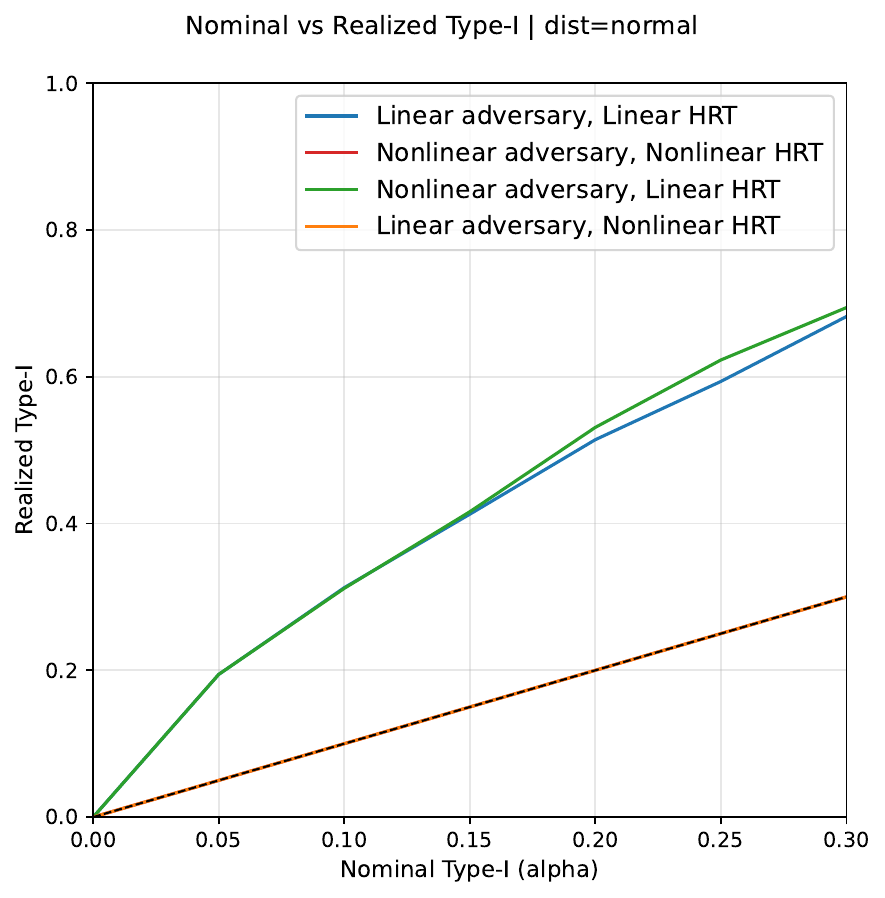}
  \caption{Normal (Type-I)}
  \label{fig:hrt_type1_norm}
\end{subfigure}\hfill

\vspace{0.6em}

\begin{subfigure}[t]{0.33\textwidth}
  \centering
  \includegraphics[width=\linewidth]{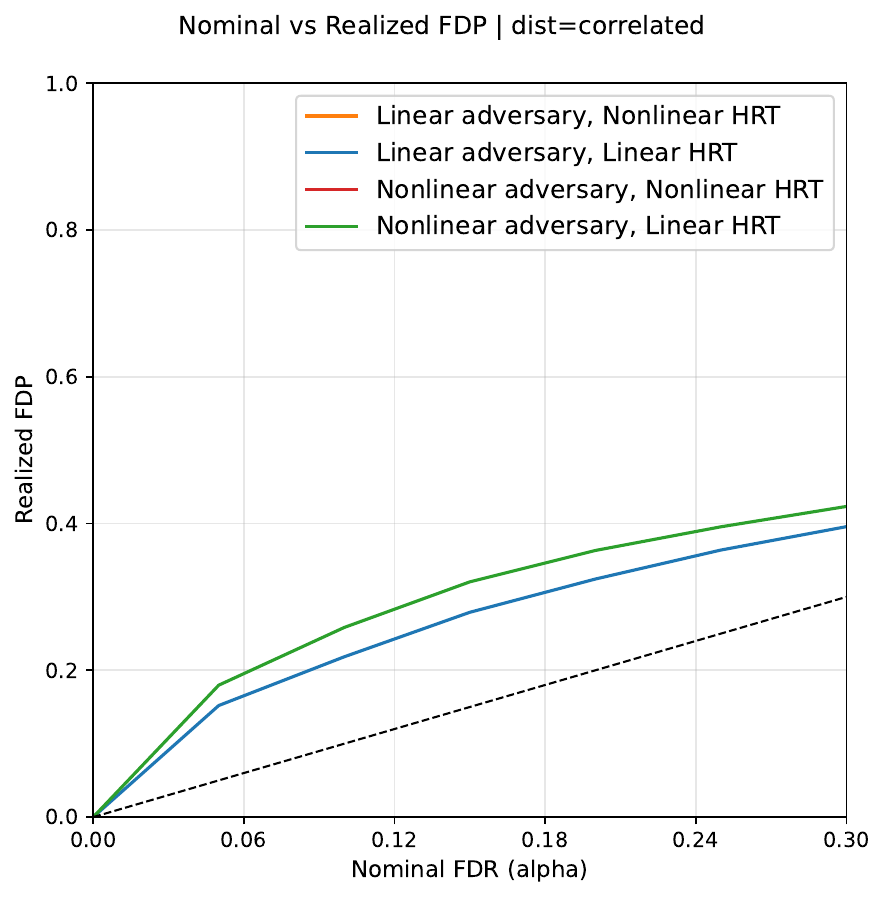}
  \caption{Correlated (FDP)}
  \label{fig:hrt_fdp_corr}
\end{subfigure}\hfill
\begin{subfigure}[t]{0.33\textwidth}
  \centering
  \includegraphics[width=\linewidth]{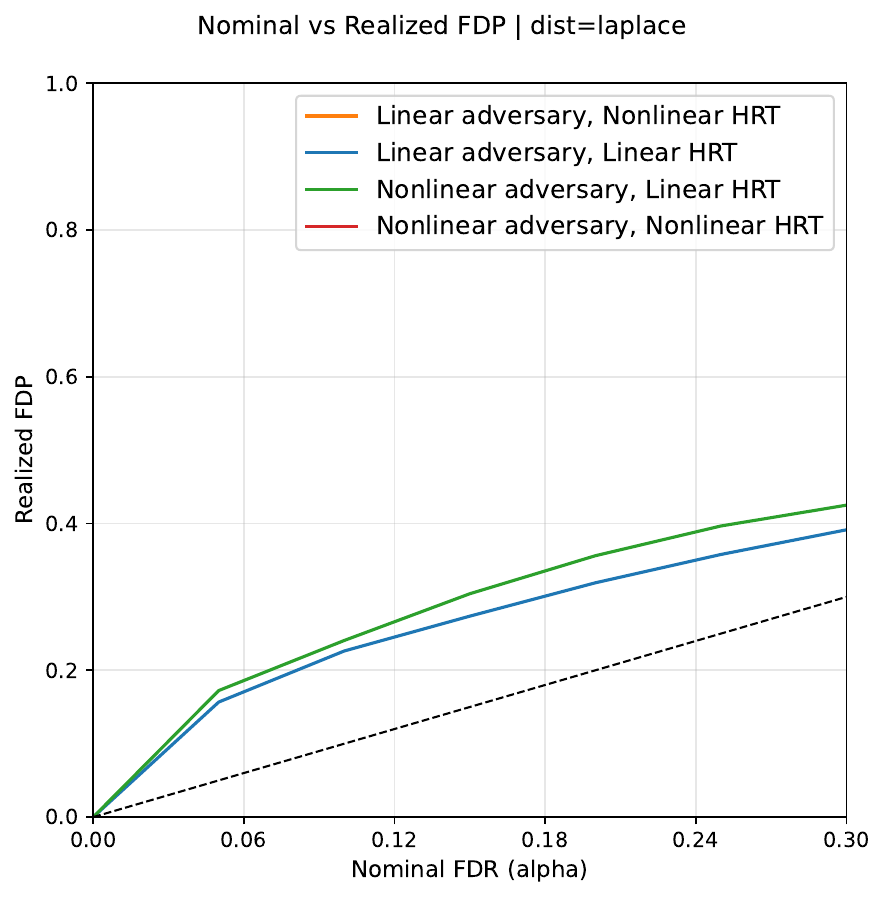}
  \caption{Laplace (FDP)}
  \label{fig:hrt_fdp_lap}
\end{subfigure}\hfill
\begin{subfigure}[t]{0.33\textwidth}
  \centering
  \includegraphics[width=\linewidth]{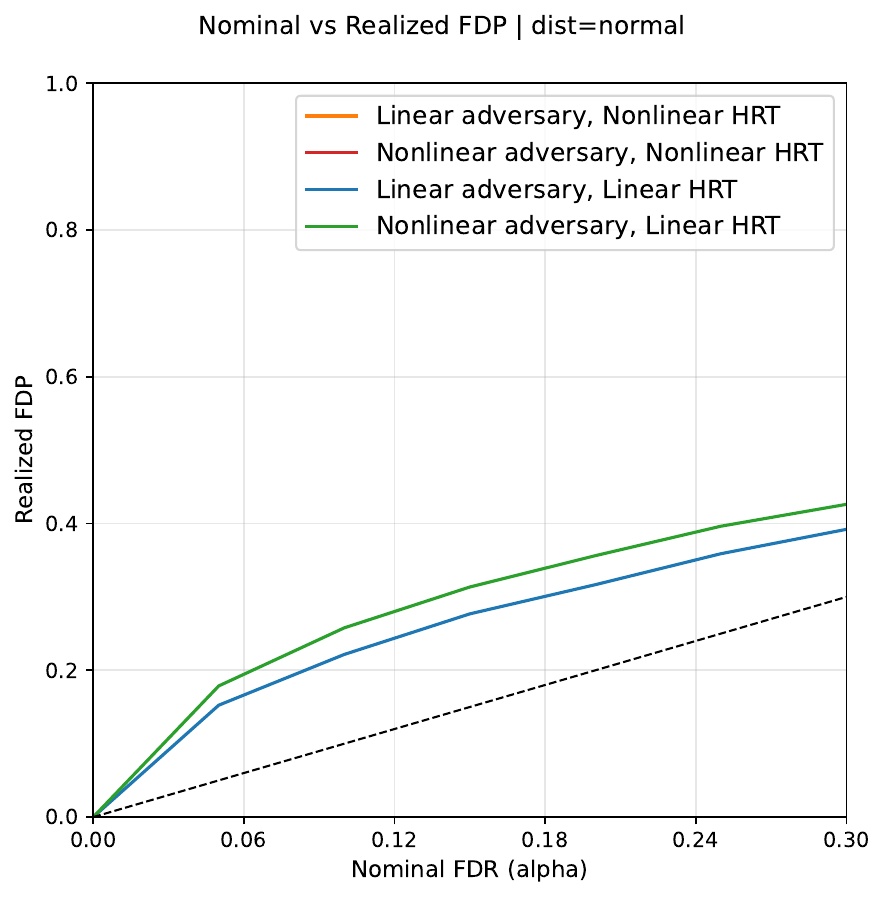}
  \caption{Normal (FDP)}
  \label{fig:hrt_fdp_norm}
\end{subfigure}\hfill

\caption{\textbf{HRT calibration across distributions.} Comparisons between different miscalibration combinations on the adversary and the test. Type-I metric (top) and FDP metric (bottom).}
\label{fig:hrt_type1_fdp_grid}
\end{figure}

\clearpage

\section{ADDITIONAL DATASETS}

We have added more experiments on two standard real world datasets widely used in other studies, the Diagnostic Wisconsin Breast Cancer dataset and the Wine recognition dataset. Both datasets are available from the UCI Machine Learning Repository. For both datasets we use exactly the same construction as in our GDSC gene expression experiments: we fix the observed covariates $X$, select a small subset of ``active'' features, and generate a nonlinear response from these actives plus noise. In the breast cancer configuration we take $n=300$ samples, $m=30$ features, and $|S_{\text{active}}| = 10$; in the wine configuration we take $n=100$, $m=13$, and $|S_{\text{active}}| = 4$. Calibration was done with our FDP metric and shown in Table \ref{tab:breast} and Table \ref{tab:wine}.

\begin{table}[t]
    \centering
    \caption{\textbf{Breast Cancer Dataset.} Entries are valid power / realized FDR.}
    \begin{tabular}{lcccc}
        \toprule
        Method & $\alpha=0.05$ & $\alpha=0.10$ & $\alpha=0.15$ & $\alpha=0.20$ \\
        \midrule
        GCM                & 0.165 / 0.086 & 0.127 / 0.141 & 0.125 / 0.181 & 0.145 / 0.227 \\
        Calibrated GCM     & 0.166 / 0.058 & 0.179 / 0.070 & 0.177 / 0.075 & 0.177 / 0.082 \\
        HRT                & 0.109 / 0.220 & 0.074 / 0.321 & 0.053 / 0.388 & 0.057 / 0.444 \\
        Calibrated HRT     & 0.161 / 0.093 & 0.146 / 0.137 & 0.137 / 0.159 & 0.150 / 0.176 \\
        CONTRA-HRT         & 0.010 / 0.489 & 0.010 / 0.497 & 0.010 / 0.502 & 0.027 / 0.520 \\
        CONTRA-FASTCRT     & 0.012 / 0.539 & 0.012 / 0.548 & 0.008 / 0.562 & 0.017 / 0.572 \\
        \bottomrule
    \end{tabular}
    \label{tab:breast}
\end{table}

\begin{table}[t]
    \centering
    \caption{\textbf{Wine Dataset.} Entries are valid power / realized FDR.}
    \begin{tabular}{lcccc}
        \toprule
        Method & $\alpha=0.05$ & $\alpha=0.10$ & $\alpha=0.15$ & $\alpha=0.20$ \\
        \midrule
        GCM                & 0.372 / 0.132 & 0.292 / 0.203 & 0.255 / 0.244 & 0.207 / 0.281 \\
        Calibrated GCM     & 0.432 / 0.043 & 0.407 / 0.085 & 0.333 / 0.158 & 0.292 / 0.204 \\
        HRT                & 0.400 / 0.030 & 0.388 / 0.050 & 0.370 / 0.076 & 0.347 / 0.103 \\
        Calibrated HRT     & 0.400 / 0.030 & 0.388 / 0.050 & 0.370 / 0.076 & 0.347 / 0.103 \\
        CONTRA-HRT         & 0.122 / 0.324 & 0.122 / 0.333 & 0.092 / 0.375 & 0.087 / 0.388 \\
        CONTRA-FASTCRT     & 0.133 / 0.335 & 0.125 / 0.352 & 0.068 / 0.405 & 0.068 / 0.425 \\
        \bottomrule
    \end{tabular}
    \label{tab:wine}
\end{table}

\vfill

\end{document}